\documentclass[%
 reprint,
superscriptaddress,
 amsmath,amssymb,
aps,
pra,
]{revtex4-2}

\usepackage{graphicx}% Include figure files
\usepackage{bm}% bold math
\usepackage{amsthm,amsmath,amssymb}
\usepackage{color}
\usepackage{enumitem}
\usepackage{caption}
\usepackage{subcaption}
\usepackage{mathtools}
\usepackage{dcolumn}% Align table columns on decimal point
\usepackage{hyperref}% add hypertext capabilities
\usepackage[ruled]{algorithm2e}
\definecolor{Red}{rgb}{1,0,0}

\def\tr{\operatorname{tr}}

\def\poly{\operatorname{poly}}
\begin{document}

\preprint{APS/123-QED}

\title{Quantum correlation alignment for unsupervised domain adaptation}

\author{Xi He}
\email{xihe@std.uestc.edu.cn}
\affiliation{Institute of Fundamental and Frontier Sciences, University of Electronic Science and Technology of China}

\begin{abstract}
Correlation alignment (CORAL), a representative domain adaptation (DA) algorithm, decorrelates and aligns a labelled source domain dataset to an unlabelled target domain dataset to minimize the domain shift such that a classifier can be applied to predict the target domain labels. In this paper, we implement the CORAL on quantum devices by two different methods. One method utilizes quantum basic linear algebra subroutines (QBLAS) to implement the CORAL with exponential speedup in the number and dimension of the given data samples. The other method is achieved through a variational hybrid quantum-classical procedure. In addition,  the numerical experiments of the CORAL with three different types of data sets, namely the synthetic data, the synthetic-Iris data, the handwritten digit data, are presented to evaluate the performance of our work. The simulation results prove that the variational quantum correlation alignment algorithm (VQCORAL) can achieve competitive performance compared with the classical CORAL.

%\begin{description}
%\end{description}
\end{abstract}


\maketitle
\section{Introduction}
\label{sec:introduction}
Quantum computation is demonstrated to have the potential to improve the performance of classical computation problems~\cite{shor1994algorithms, grover1996fast, harrow2009quantum, aaronson2011computational, farhi2018classification, arute2019quantum}. In addition, quantum computation can be applied to accomplish machine learning tasks with quantum speedup~\cite{lloyd2013quantum, havlivcek2019supervised, schuld2019quantum, schuld2018supervised}. Originally, many quantum shallow machine learning algorithms are proposed such as quantum principal component analysis~\cite{lloyd2014quantum}, quantum classification~\cite{rebentrost2014quantum, schuld2020circuit, schuld2017implementing, schuld2014quantum, schuld2018quantum}, quantum data fitting~\cite{wiebe2012quantum, schuld2016prediction}, quantum clustering~\cite{aimeur2013quantum, wiebe2018quantum} and quantum dimensionality reduction~\cite{cong2016quantum, duan2019quantum, he2019quantumLLE}. In recent years, quantum auto-encoders~\cite{romero2017quantum}, quantum Boltzmann machine~\cite{wiebe2014quantum, amin2018quantum}, quantum generative adversarial network~\cite{lloyd2018quantum, dallaire2018quantum} and quantum feedforward neural network~\cite{wan2017quantum} are the representative quantum deep learning models. For transfer learning, a significant research subfield of machine learning, it can also be combined with quantum computation to implement machine learning tasks in a different, but related domain with the acquired knowledge of a well-studied domain~\cite{mari2019transfer, he2019quantumTCA}.

In the field of machine learning, labelled data sets are actually dreadfully scarce compared with the available huge amount of unlabelled data. In most cases, the collected unprocessed data are labelled by the extremely time-consuming manual labeling method. Domain adaptation (DA), a crucial research branch of transfer learning, aims to predict the labels of an unprocessed target domain dataset with a labelled source domain dataset~\cite{pan2009survey}. It has various applications in computer vision~\cite{csurka2017domain}, natural language processing~\cite{glorot2011domain} and reinforcement learning~\cite{carr2019domain}. It can be mainly categorized into the semi-supervised DA, few labels in the target domain, and the unsupervised DA, no labels available in the target domain. For the unsupervised DA, the data distribution adaptation~\cite{pan2011domain, long2013transfer, wang2017balanced} which attempts to approximate the data distributions of the source and target domain datasets is one of the most representative domain adaptation methods. In addition, the subspace projection~\cite{fernando2013unsupervised, gopalan2011domain, gong2012geodesic} is another common method for the DA. It firstly projects the original given data to a specified subspace and subsequently reduces the domain shift by aligning the subspaces. Different from the two methods above, the correlation alignment algorithm (CORAL)~\cite{sun2016return, sun2016correlation} is a simpler but efficient DA algorithm.

The CORAL firstly decorrelates the labelled source domain data to eliminate its unique data characteristics. Subsequently, it aligns the decorrelated labelled source domain data to the unlabelled target domain data to reduce the domain shift. The goal of the CORAL is to minimize the discrepancy between the source and target domain datasets by aligning their second order statistics, namely the covariance matrices~\cite{sun2016return}. The CORAL directly aligns the datasets without projecting the data to their corresponding subspaces resulting in a much more concise procedure than other DA methods. After the data decorrelation and alignment, a classifier will be trained on the aligned labelled source domain dataset and applied to the unlabelled target domain dataset to predict the target domain labels. With the CORAL, the labels of an unprocessed target domain can be obtained efficiently without the need for the costly manual labeling. However, the algorithmic complexity of the CORAL can be prohibited with the increase of the number and dimension of the given data.

In our work, two different types of quantum implementations of the CORAL are presented. One implementation, namely the QBLAS-based CORAL, can be performed on a universal quantum computer achieving exponential speedup in the number and dimension of the given data. The other implementation, the VQCORAL can be performed on the near-term quantum devices through a variational hybrid quantum-classical procedure. Concretely, the VQCORAL can be realized in two different ways called the end-to-end VQCORAL and the matrix-multiplication-based VQCORAL which is inspired from the variational quantum eigensolver (VQE)~\cite{peruzzo2014variational, higgott2019variational} and the variational quantum matrix multiplication~\cite{bravo2019variational}. To evaluate the performance of the VQCORAL, three different numerical experiments are provided. Specifically, the no adaptation model (NA) set as the baseline model, the classical CORAL, the VQCORAL are the models selected in the experiments. For the two synthetic data sets generated from different distributions, the VQCORAL outperforms the classical CORAL and the NA with a two-qubit eight-layer variational quantum circuit. For the synthetic-Iris data sets~\cite{fisher1936use, anderson1936species}, the VQCORAL also shows outstanding performance with a two-qubit eight-layer parameterized quantum circuit compared to the other two models. For the handwritten digit datasets, namely the MNIST~\cite{lecun1998gradient} and USPS~\cite{lecun1990handwritten} data sets, the DA procedure can be implemented by an eight-qubit sixteen-layer parameterized quantum circuit to achieve comparable performance to the classical CORAL and better than the baseline model.

The arrangement of this paper is shown as follows. In section~\ref{sec:classical CORAL}, the classical CORAL will be briefly overviewed. Subsequently, the quantum correlation alignment (QCORAL) is presented in section~\ref{sec:quantum CORAL}. The QBLAS-based CORAL and the VQCORAL are shown in section~\ref{subsec:QBLAS-based CORAL} and section~\ref{subsec:VQCORAL} respectively in detail. Then, the numerical experiments are provided in section~\ref{sec:numerical experiments}. Finally, we make a conclusion and discuss some open questions in section~\ref{sec:discussions}. 

\section{Classical correlation alignment}
\label{sec:classical CORAL}
Given a labelled source domain dataset $D_{s} = \{ x_{i}^{(s)} \}_{i=1}^{n_{s}} \in \mathbb{R}^{D}$ with labels $L_{s} = \{ y_{i}^{(s)} \}_{i=1}^{n_{s}}$ and an unlabelled target domain dataset $D_{t} = \{ x_{j}^{(t)} \}_{j=1}^{n_{t}} \in \mathbb{R}^{D}$ generated from different data distributions. $X_{s} = (x_{1}^{(s)}, \dots, x_{n_{s}}^{(s)}) \in \mathbb{R}^{D \times n_{s}}$, $X_{t} = (x_{1}^{(t)}, \dots, x_{n_{t}}^{(t)}) \in \mathbb{R}^{D \times n_{t}}$ refer to the source and target domain dataset matrices respectively. Assume $u_{s}$ ($u_{t}$), $C_{s}$ ($C_{t}$) are the mean and covariance matrix of the source (target) domain respectively. The data in both domains have been zero-centered, namely $u_{s} = u_{t} = 0$, and normalized but $C_{s} \neq C_{t}$. In addition, the data in the CORAL are assumed to depend on a lower-dimensional manifold, meaning that $X_{s}$, $X_{t}$, $C_{s}$, $C_{t}$ are all low-rank matrices where $r_{C_{s}}$, $r_{C_{t}}$ represents the rank of $C_{s}$, $C_{t}$ respectively. 

\begin{figure}
	\centering
	\includegraphics[width=\columnwidth]{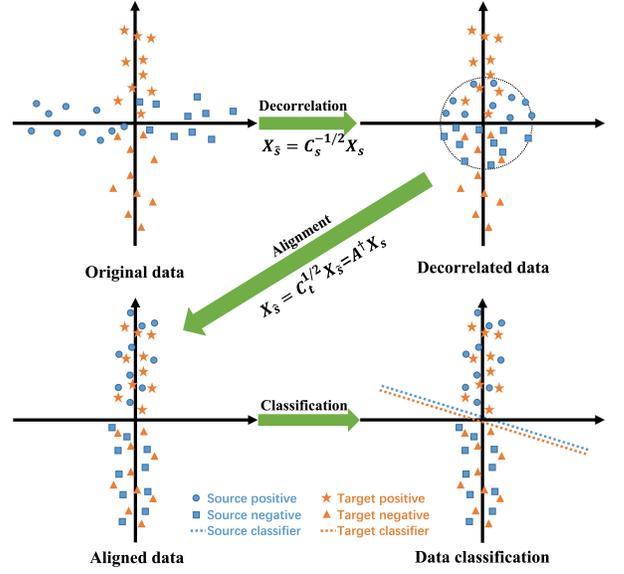}	
	\caption{The schematic diagram of the CORAL}
	\label{fig:CORAL}
\end{figure} 

The CORAL attempts to align the covariance matrix of the source domain to the target domain utilizing a linear transformation matrix $A$~\cite{sun2016return}. Thus, the objective function of the CORAL is defined as
\begin{equation}
	\min_{A} \Vert C_{\hat{s}} - C_{t} \Vert_{F}^{2} = \min_{A} \Vert A^{T} C_{s} A - C_{t} \Vert_{F}^{2},
	\label{eq:objective function}
\end{equation}
where $C_{\hat{s}} = A^{T} C_{s} A$ is the covariance matrix after the correlation alignment; $\Vert \cdot \Vert_{F}$ represents the Frobenius norm. 

Assume $C_{s} = U_{s} \Sigma_{s} U_{s}^{T}$, $C_{t} = U_{t} \Sigma_{t} U_{t}^{T}$ are the singular value decomposition (SVD) of $C_{s}$, $C_{t}$ respectively. The optimal solution of Eq.~\eqref{eq:objective function} is $C_{\hat{s}} = U_{t[1:r]} \Sigma_{t[1:r]} U_{t[1:r]}^{T}$ where $r = \min(r_{C_{s}}, r_{C_{t}})$; the diagonal elements of $\Sigma_{t[1:r]}$ are the $r$ largest singular values; the columns of $U_{t[1:r]}$ are the corresponding left-singular vectors. Let
\begin{equation}
	C_{\hat{s}} = A^{T} C_{s} A = U_{t[1:r]} \Sigma_{t[1:r]} U_{t[1:r]}^{T}.
	\label{eq:covariance matrix} 
\end{equation}
Then 
\begin{equation}
	A^{T} U_{s} \Sigma_{s} U_{s}^{T} A = U_{t[1:r]} \Sigma_{t[1:r]} U_{t[1:r]}^{T}.
	\label{eq:subtraction singular value}
\end{equation}
Hence,
\begin{equation}
	(U_{s}^{T} A)^{T} \Sigma_{s} (U_{s}^{T} A) = E^{T} \Sigma_{s} E
	\label{eq:transformation of singular value equation}
\end{equation}
where $E = \Sigma_{s}^{+ \frac{1}{2}} U_{s}^{T} U_{t[1:r]} \Sigma_{t[1:r]}^{\frac{1}{2}} U_{t[1:r]}^{T}$; $\Sigma_{s}^{+ \frac{1}{2}}$ is the Moore-Penrose pseudoinverse of $\Sigma_{s}^{\frac{1}{2}}$.

Finally, the optimal solution of $A$ is 
\begin{align}
	A_{\ast} &= U_{s} E \notag \\ 
	&= (U_{s}\Sigma_{s}^{+ \frac{1}{2}} U_{s}^{T}) (U_{t[1:r]} \Sigma_{t[1:r]}^{\frac{1}{2}} U_{t[1:r]}^{T}).
	\label{eq:equation transformation}
\end{align}
The first term $U_{s}\Sigma_{s}^{+ \frac{1}{2}} U_{s}^{T}$ decorrelates the source domain dataset. The second term $U_{t[1:r]} \Sigma_{t[1:r]}^{\frac{1}{2}} U_{t[1:r]}^{T}$ aligns the decorrelated source domain dataset to the target domain dataset.  

Therefore, the concrete steps of the CORAL are as follows:

(1) Compute the source domain covariance matrix $C_{s} = X_{s}X_{s}^{T}$ and the target domain covariance matrix $C_{t} = X_{t}X_{t}^{T}$.

(2) Decorrelate the source domain data as
\begin{equation}
	X_{\tilde{s}} = C_{s}^{-\frac{1}{2}} X_{s}.
	\label{eq:decorrelated X_s}
\end{equation}

(3) Align the decorrelated source domain data to the target domain data as
\begin{equation}
	X_{\hat{s}} = C_{t}^{\frac{1}{2}} X_{\tilde{s}}.
	\label{eq:aligned X_s}
\end{equation}

After the CORAL, the source domain data are transformed to the target domain data space. The classifier can be subsequently trained on the aligned source domain data $\{ x_{i}^{(\hat{s})}, y_{i}^{(s)} \}_{i=1}^{n_{s}}$ and predict the labels $L_{t} = \{ y_{j}^{(t)} \}_{j=1}^{n_{t}}$ of the target domain data $D_{t}$. The schematic diagram of the CORAL is presented in Fig.~\ref{fig:CORAL}. 

\section{Quantum correlation alignment}
\label{sec:quantum CORAL}
The quantum correlation alignment algorithm (QCORAL) can be implemented in two aspects, based on the quantum basic linear algebra subroutines and the variational hybrid quantum-classical procedure respectively. In these two implementations, we assume that all the data have been normalized and zero-centered exactly as the classical CORAL. 

\subsection{State preparation}
\label{subsec:state preparation}
Given the source domain data $X_{s} = \sum_{i=1}^{n_s} \vert x_{i}^{(s)} \vert | x_{i}^{(s)} \rangle \langle i |$ and the target domain data $X_{t} = \sum_{j=1}^{n_t} \vert x_{j}^{(t)} \vert | x_{j}^{(t)} \rangle \langle j |$. The quantum states representing the source domain data $X_{s}$ and the target domain data $X_{t}$ are 
\begin{equation}
	| \psi_{X_s} \rangle = \sum_{i=1}^{n_s} \sum_{m=1}^{D} x_{mi}^{(s)} | i \rangle | m \rangle = \sum_{i=1}^{n_{s}} | i \rangle | x_{i}^{(s)} \rangle,
	\label{eq:source domain state}
\end{equation}
\begin{equation}
	| \psi_{X_t} \rangle = \sum_{j=1}^{n_t} \sum_{m=1}^{D} x_{mj}^{(t)} | j \rangle | m \rangle = \sum_{j=1}^{n_{t}} | j \rangle | x_{j}^{(t)} \rangle,
	\label{eq:target domain state}
\end{equation}
respectively in amplitude encoding with $\sum_{m, i} \vert x_{mi}^{(s)} \vert = \sum_{m, j} \vert x_{mj}^{(t)} \vert = 1$. Hence, the covariance matrices of the source and target domain data can be obtained as
\begin{align}
	\rho_{C_{s}} &= \tr_{i}\{ | \psi_{X_s} \rangle \langle \psi_{X_s} | \} \notag \\
	&= \sum_{m, m^{'} = 1}^{D} \sum_{i=1}^{n_s} x_{mi}^{(s)} x_{m^{'}i}^{(s) \ast} | m \rangle \langle m^{'} |,
	\label{eq:source domain covariance}
\end{align}
\begin{align}
	\rho_{C_{t}} &= \tr_{j}\{ | \psi_{X_t} \rangle \langle \psi_{X_t} | \} \notag \\
	&= \sum_{m, m^{'} = 1}^{D} \sum_{j=1}^{n_t} x_{mj}^{(t)} x_{m^{'}j}^{(t) \ast} | m \rangle \langle m^{'} |,
	\label{eq:target domain covariance}
\end{align}
respectively by taking the partial trace over the corresponding column register.

\subsection{QBLAS-based CORAL}
\label{subsec:QBLAS-based CORAL}
The QBLAS-based CORAL utilizes the quantum basic linear algebra subroutines to implement the data decorrelation and alignment procedure of the CORAL. In the spirit of~\cite{rebentrost2018quantum}, the source domain data $X_{s}$ can be aligned to the target domain data $X_{t}$ as follows. 

Assume the elements of $X_{s}$ and $X_{t}$ are accessible in a quantum random access memory~\cite{giovannetti2008quantum}. Let $X_{s} = \sum_{m} \sigma_{m}^{(s)} | u_{m}^{(s)} \rangle \langle v_{m}^{(s)} |$, $X_{t} = \sum_{m} \sigma_{m}^{(t)} | u_{m}^{(t)} \rangle \langle v_{m}^{(t)} |$ be the SVD of $X_{s}$ and $X_{t}$ respectively. The source and target domain data $X_{s}$, $X_{t}$ can be extended to  
\begin{equation}
	\tilde{X}_{s} = \begin{bmatrix}
		0 & X_{s} \\
		X_{s}^{\dagger} & 0
	\end{bmatrix},
	\label{eq:extended source domain matrix}
\end{equation}
\begin{equation}
	\tilde{X}_{t} = \begin{bmatrix}
		0 & X_{t} \\
		X_{t}^{\dagger} & 0 
	\end{bmatrix}.
	\label{eq:extended target domain matrix}
\end{equation}

With the input state $| 0, \psi_{X_s} \rangle | 0 \rangle^{\otimes \log(D + n_s)}$, the quantum state 
\begin{align}
	&\sum_{i=1}^{n_{s}} | i \rangle \sum_{m=1}^{D} \beta_{mi}^{(s)} | \sigma_{m}^{(s)} \rangle \frac{1}{\sqrt{2}} (| w_{m}^{(s)+}\rangle - | w_{m}^{(s)-} \rangle) \notag \\ 
	&= \sum_{i=1}^{n_{s}} | i \rangle \sum_{m=1}^{D} \beta_{mi}^{(s)} | \sigma_{m}^{(s)} \rangle | v_{m}^{(s)} \rangle 
	\label{eq:PE source state}
\end{align}
can be obtained by performing the quantum phase estimation (QPE) 
\begin{align}
	\textbf{U}_{\textbf{PE}}(\tilde{X}_{s}) = &(\textbf{QFT}^{\dagger} \otimes \textbf{I}) \left( \sum_{\tau=0}^{T-1} | \tau \rangle \langle \tau | \otimes e^{i \tilde{X}_{s} \tau t / T} \right) \notag \\
	&(\textbf{H}^{\otimes n} \otimes \textbf{I})
	\label{eq:phase estimation}
\end{align}
as described in~\cite{harrow2009quantum, duan2017quantum} where $\beta_{mi}^{(s)} = \langle u_{m}^{(s)} | x_{i}^{(s)} \rangle$; $| w_{m}^{(s) \pm} \rangle = \frac{1}{\sqrt{2}} (| 0 \rangle | u_{m}^{(s)} \rangle \pm | 1 \rangle | v_{m}^{(s)} \rangle)$ are the eigenvectors of $\tilde{X}_{s}$ corresponding to the singular value $\sigma_{m}^{(s)}$; $\textbf{QFT}^{\dagger}$ represents the inverse quantum Fourier transform and $\sum_{\tau=0}^{T-1} | \tau \rangle \langle \tau | \otimes e^{i \tilde{X}_{s} \tau t / T}$ is the conditional Hamiltonian evolution. Subsequently, add a new ancilla qubit and apply the rotation operation $R_{y}(\sin^{-1} (\gamma_{s} / \vert \sigma_{m}^{(s)} \vert))$ on it resulting in 
\begin{equation}
	\sum_{i=1}^{n_{s}} | i \rangle \sum_{m=1}^{D} \beta_{mi}^{(s)} | \sigma_{m}^{(s)} \rangle | v_{m}^{(s)} \rangle | \psi_{a}^{(s)} \rangle
	\label{eq:rotation source state}
\end{equation}
where the ancilla register
\begin{equation}
	| \psi_{a}^{(s)} \rangle = \sqrt{1 - \frac{\gamma_{s}^{2}}{\vert \sigma_{m}^{(s)} \vert^{2}}} | 0 \rangle + \frac{\gamma_{s}}{\vert \sigma_{m}^{(s)} \vert} | 1 \rangle,
	\label{eq:psi_s_a}
\end{equation}
$\gamma_{s}$ is a constant.
By uncomputing the singular value register and measure the ancilla register to be $| 1 \rangle$, the decorrelated source domain quantum state
\begin{align}
	| \psi_{X_{\tilde{s}}} \rangle &= \sum_{i=1}^{n_{s}} |i \rangle \sqrt{\frac{1}{\sum_{m=1}^{D} \vert \gamma_{s} \beta_{mi}^{(s)} \vert^{2} / \vert \sigma_{m}^{(s)} \vert^2}}\sum_{m=1}^{D} \frac{\beta_{mi}^{(s)} \gamma_{s}}{\vert \sigma_{m}^{(s)} \vert} | v_{m}^{(s)} \rangle \notag \\ 
	&= \sum_{i=1}^{n_{s}} | i \rangle \frac{C_{s}^{-\frac{1}{2}} | x_{i}^{(s)} \rangle} {\tr(C_{s}^{-\frac{1}{2}} | x_{i}^{(s)} \rangle)} \notag \\
	&= \sum_{i=1}^{n_{s}} | i \rangle | x_{i}^{(\tilde{s})} \rangle
	\label{eq:decorrelated quantum source state} 
\end{align}
representing the decorrelated source domain dataset $X_{\tilde{s}}$ is finally obtained. Hence, the source domain data can be decorrelated in $O(\Vert X_{s} \Vert_{\max}^{2} \log^2(D + n_{s}) / \epsilon^{3})$ where $\Vert X_{s} \Vert_{\max}$ is the largest absolute element of $X_{s}$ and $\epsilon$ is the error parameter~\cite{rebentrost2018quantum}.  

Similarly, we then perform the QPE $\textbf{U}_{\textbf{PE}}(\tilde{X}_{t})$ on $| 0, \psi_{X_{\tilde{s}}} \rangle | 0 \rangle^{\otimes \log(D+n_t)}$ resulting in 
\begin{align}
	&\sum_{i=1}^{n_{s}} | i \rangle \sum_{m=1}^{D} \beta_{mi}^{(t)} | \sigma_{m}^{(t)} \rangle \frac{1}{\sqrt{2}} (| w_{m}^{(t)+}\rangle - | w_{m}^{(t)-} \rangle) \notag \\ 
	&= \sum_{i=1}^{n_{s}} | i \rangle \sum_{m=1}^{D} \beta_{mi}^{(t)} | \sigma_{m}^{(t)} \rangle | v_{m}^{(t)} \rangle,
	\label{eq:PE target state} 
\end{align}
where $\beta_{mi}^{(t)} = \langle u_{m}^{(t)} | x_{i}^{(\tilde{s})} \rangle$; $| w_{m}^{(t) \pm} \rangle = \frac{1}{\sqrt{2}} (| 0 \rangle | u_{m}^{(t)} \rangle \pm | 1 \rangle | v_{m}^{(t)} \rangle)$ are the eigenvectors of $\tilde{X}_{t}$ corresponding to the singular value $\sigma_{m}^{(t)}$. By performing the rotation operation $R_{y}(\sin^{-1} (\gamma_{t} \vert \sigma_{m}^{(t)} \vert))$ on a newly added ancilla, the quantum state
\begin{equation}
	\sum_{i=1}^{n_{s}} | i \rangle \sum_{m=1}^{D} \beta_{mi}^{(t)} | \sigma_{m}^{(t)} \rangle | v_{m}^{(t)} \rangle \left( \sqrt{1 - \gamma_{t}^{2} \vert \sigma_{m}^{(t)} \vert^{2}} | 0 \rangle + \gamma_{t} \vert \sigma_{m}^{(t)} \vert | 1 \rangle \right)
	\label{eq:rotation target state}
\end{equation}
is achieved where $\gamma_{t}$ is a constant.
Ultimately, the quantum state
\begin{align}
	| \psi_{X_{\hat{s}}} \rangle &= \sum_{i=1}^{n_{s}} | i \rangle \sqrt{\frac{1}{\sum_{m=1}^{D} \vert \gamma_{t} \beta_{mi}^{(t)} \sigma_{m}^{(t)} \vert^{2}}}\sum_{m=1}^{D} \beta_{mi}^{(t)} \gamma_{t} \vert \sigma_{m}^{(t)} \vert | v_{m}^{(t)} \rangle \notag \\
	&= \sum_{i=1}^{n_{s}} | i \rangle \frac{C_{t}^{\frac{1}{2}} | x_{i}^{(\tilde{s})} \rangle} {\tr(C_{t}^{\frac{1}{2}} | x_{i}^{(\tilde{s})} \rangle)} \notag \\
	&= \sum_{i=1}^{n_{s}} | i \rangle | x_{i}^{(\hat{s})} \rangle
	\label{eq:aligned quantum source state} 
\end{align}
can be obtained in $O(\Vert X_{t} \Vert_{\max}^{2} \log^{2} (D + n_{t}) / \epsilon^{3})$ where $\Vert X_{t} \Vert$ is the largest absolute element of $X_{t}$~\cite{rebentrost2018quantum}. Therefore, the decorrelated source domain data are aligned to the target domain data. After the data decorrelation and alignment, the classifier is applied to the aligned source domain data $X_{\hat{s}}$ with labels $L_{s}$ and the target domain data $X_{t}$ to predict the target labels $L_{t}$. The pseudo-code of the QBLAS-based CORAL is presented in Algorithm~\ref{alg:QBLAS-based CORAL}. In contrast, the implementation of the classical CORAL involves SVD and matrix multiplication operations resulting in the algorithmic complexity in $O(\poly(n_{s}, n_{t}, D))$. Thus, the QBLAS-based CORAL presented in this subsection takes logarithmic resources in the number and dimension of the source and target domain data compared to the classical CORAL.

\begin{algorithm}[htb]
	\caption{QBLAS-based CORAL}
	%\LinesNumbered
	\KwIn{Source domain data $X_{s}$ with labels $L_{s}$; target domain data $X_{t}$.}
	\KwOut{Target domain labels $L_{t}$.}
	\emph{step 1}: Apply the QPE $\textbf{U}_{\textbf{PE}}(\tilde{X}_{s})$ on the input state $| 0, \psi_{X_s} \rangle | 0 \rangle^{\otimes \log(D + n_s)}$ resulting in Eq.~\eqref{eq:PE source state} in $O(\frac{1}{\epsilon})$ with an error $\epsilon$. \\
	\emph{step 2}: Add a new ancilla and perform the rotation operation $R_{y}(\sin^{-1} (\gamma_{s} / \vert \sigma_{m}^{(s)} \vert))$ to obtain Eq.~\eqref{eq:rotation source state}. \\
	\emph{step 3}: Uncompute the singular value register $| \sigma_{m}^{(s)} \rangle$ and measure the ancilla register to be $| 1 \rangle$ to obtain the decorrelated source domain quantum state $| \psi_{X_{\tilde{s}}} \rangle$ as Eq.~\eqref{eq:decorrelated quantum source state} in $O(\Vert X_{s} \Vert_{\max}^{2} \log^2(D + n_{s}) / \epsilon^{3})$. \\
	\emph{step 4}: Perform $\textbf{U}_{\textbf{PE}}(\tilde{X}_{t})$ on $| 0, \psi_{X_{\tilde{s}}} \rangle | 0 \rangle^{\otimes \log(D+n_t)}$ resulting in the quantum state as Eq.~\eqref{eq:PE target state}. \\
	\emph{step 5}: Perform the rotation operation $R_{y}(\sin^{-1}(\gamma_{t} \vert \sigma^{(t)}_{m} \vert))$ on a newly added ancilla to obtain Eq.~\eqref{eq:rotation target state}. \\
	\emph{step 6}: Uncompute the singular value register $| \sigma_{m}^{(t)} \rangle$ and measure the ancilla to be $| 1 \rangle$ to achieve the aligned source domain quantum state $| \psi_{X_{\hat{s}}} \rangle$ as Eq.~\eqref{eq:aligned quantum source state} in $O(\Vert X_{t} \Vert_{\max}^{2} \log^{2} (D + n_{t}) / \epsilon^{3})$. \\
	\emph{step 7}: Invoke a classifier to predict the target labels $L_{t} = Classifier(X_{\hat{s}}, L_{s}, X_{t})$.
	\label{alg:QBLAS-based CORAL}
\end{algorithm}

\subsection{Variational quantum correlation alignment}
\label{subsec:VQCORAL}
Although the QBLAS-based CORAL can be performed on a universal quantum computer with exponential speedup, the implementation critically requires a high-depth quantum circuit and fully coherent evolution. Alternatively, the CORAL can be implemented on the near-term noisy intermediate-scale quantum devices with a variational hybrid quantum-classical procedure. The VQCORAL combines the quantum computation and classical optimization together to implement the algorithm with low-depth quantum circuits. In this section, we will present the implementation of the VQCORAL and explore two different specific configurations in detail.

\begin{figure}
	\centering
	\includegraphics[width=\columnwidth]{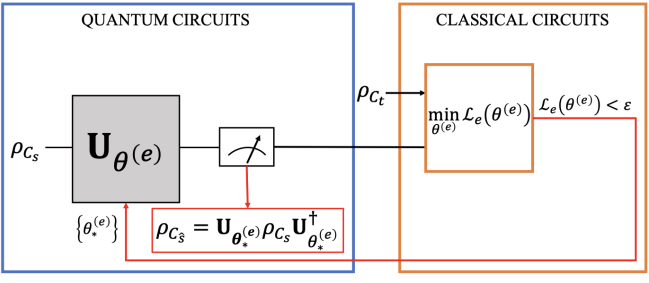}	
	\caption{The schematic diagram of the end-to-end VQCORAL}
	\label{fig:end-to-end VQCORAL}
\end{figure}

As introduced in section~\ref{sec:classical CORAL}, the goal of the CORAL is to find a linear transformation matrix $A$ to align the source domain data $X_{s}$ to the target domain data $X_{t}$. Hence, we can approximate the linear transformation by a parameterized quantum circuit $\textbf{U}_{\theta}$. The cost function of the VQCORAL can be defined as 
\begin{equation}
	L_{v}(\theta) = \Vert \textbf{U}_{\theta} \rho_{C_{s}} \textbf{U}_{\theta}^{\dagger} - \rho_{C_{t}} \Vert_{F}^{2} 
	\label{eq:VQCORAL cost function}
\end{equation}
where 
\begin{equation}
	\textbf{U}_{\theta} = \textbf{U}_{L}(\theta) \cdots \textbf{U}_{l}(\theta) \cdots \textbf{U}_{1}(\theta)
	\label{eq:U_theta}
\end{equation}
is an $L$-depth parameterized quantum circuit with a set of parameter $\{\theta\}$. Then, the optimal configuration of the quantum circuit can be obtained by minimizing $L_{v}$ with the optimization algorithm. Inspired by the classical neural network, this procedure can be called the end-to-end VQCORAL, and the corresponding schematic diagram is shown as Fig.~\ref{fig:end-to-end VQCORAL}. 

In addition to the end-to-end VQCORAL described as above, the matrix-multiplication-based VQCORAL can also be implemented in two variational procedures successively as follows:

(1) We do not optimize the cost function $L_{v}$ directly, but compute $C_{s}^{1 / 2}$ and $C_{t}^{1 / 2}$ by solving the eigenvalues and corresponding eigenvectors of $C_{s}$ and $C_{t}$ respectively by the covariance matrix square root solver (VQCMSR) inspired from Ref.~\cite{peruzzo2014variational, higgott2019variational} as presented in Algorithm~\ref{alg:VQCMSR} and depicted in Fig.~\ref{VQCMSR}.

\begin{figure*}
	\centering
	\includegraphics[width=\textwidth]{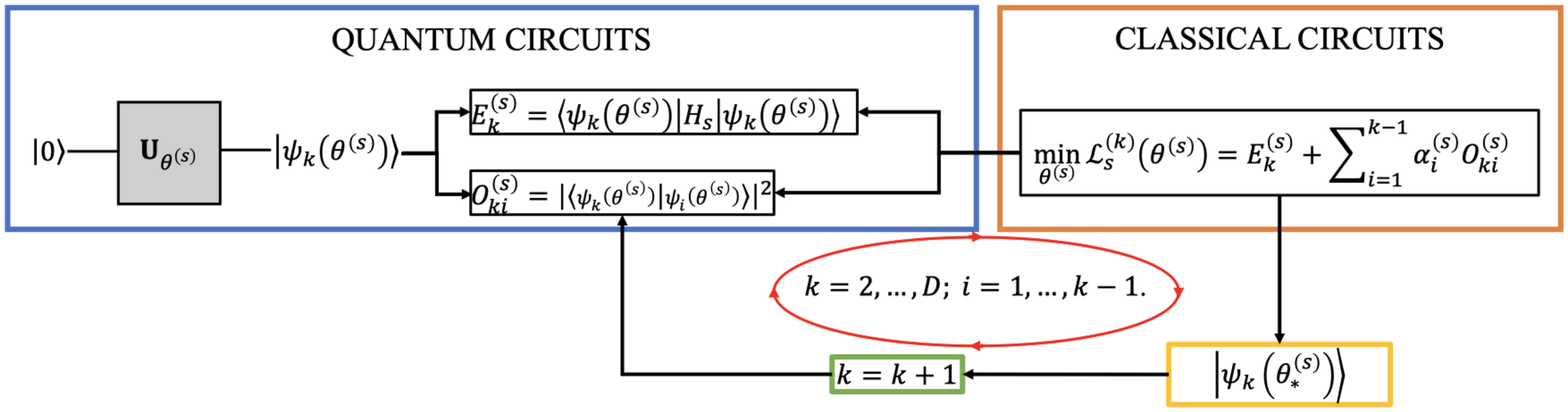}
	\includegraphics[width=\textwidth]{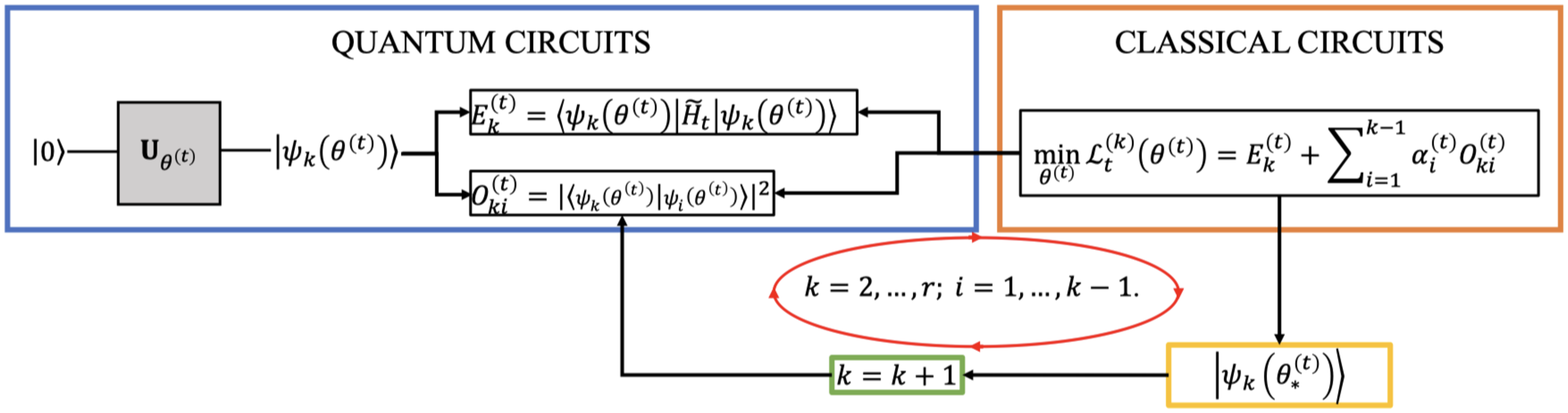}
	\caption{The schematic diagram of the VQCMSR}
	\label{fig:VQCMSR}
\end{figure*} 

\begin{algorithm}[htb]
	\caption{Variational quantum covariance matrix square root solver}
	%\LinesNumbered
	\KwIn{Source domain data $X_{s}$ with labels $L_{s}$; target domain data $X_{t}$.}
	\KwOut{The source domain covariance square root matrix $C_{s}^{1 / 2}$ and the target domain covariance square root matrix $C_{t}^{1 / 2}$.}
	\emph{step 1}: Compute the Hamiltonian $H_{s} = \rho_{C_{s}}$, $H_{t} = \rho_{C_{t}}$ and subsequently $\tilde{H}_{t} = \eta I - H_{t}$ with a specified constant $\eta$. \\
	\emph{step 2}: Prepare the ansatz states $| \psi(\lambda_{k}^{(s)}) \rangle$ with a set of parameters $\{ \theta^{(s)} \}$. Minimize the cost function 
	\begin{equation}
		F_{s}(\lambda_{k}^{(s)}) = \begin{cases}
			E_{1}^{(s)}, &k=1, \\
			E_{k}^{(s)} + \sum_{i=1}^{k-1} \alpha_{i}^{(s)} O_{ki}^{(s)}, &k=2, \cdots, D, \notag
		\end{cases}
		\label{eq:source domain vqe cost function}
	\end{equation}
	to obtain the $D$ eigenvalues of $H_{s}$ and the corresponding eigenvectors where 
	\begin{equation}
		\begin{cases}
			E_{k}^{(s)} = \langle \psi(\lambda_{k}^{(s)}) | H_{s} | \psi(\lambda_{k}^{(s)}) \rangle \notag \\
			O_{ki}^{(s)} = | \langle \psi(\lambda_{k}^{(s)}) | \psi(\lambda_{i}^{(s)}) \rangle |^{2}
		\end{cases}	
	\end{equation}
	with the weight coefficient $\alpha_{i}^{(s)}$ for $i = 1, \cdots, k-1$. \\
	\emph{step 3}: Prepare the ansatz states $| \psi(\lambda_{k}^{(t)}) \rangle$ with a set of parameters $\{ \theta^{(t)} \}$. Minimize the cost function 
	\begin{equation}
		F_{t}(\lambda_{k}^{(t)}) = \begin{cases}
			E_{1}^{(t)}, &k=1, \\
			E_{k}^{(t)} + \sum_{i=1}^{k-1} \alpha_{i}^{(t)} O_{ki}^{(t)}, &k=2, \cdots, r,
		\end{cases} \notag
	\end{equation}
	to obtain the $r$ smallest eigenvalues of $\tilde{H}_{t}$ and the corresponding eigenvectors where 
	\begin{equation}
		\begin{cases}
			E_{k}^{(t)} = \langle \psi(\lambda_{k}^{(t)}) | \tilde{H}_{t} | \psi(\lambda_{k}^{(t)}) \rangle \notag \\
			O_{ki}^{(t)} = | \langle \psi(\lambda_{k}^{(t)}) | \psi(\lambda_{i}^{(t)}) \rangle |^{2}
		\end{cases}
	\end{equation}
	with the weight coefficient $\alpha_{i}^{(t)}$ for $i = 1, \cdots, k-1$. \\
	\emph{step 4}: Compute $C_{s}^{1 / 2} = U_{s} \Sigma_{s}^{1 / 2} U_{s}^{T}$ and $C_{t}^{1 / 2} = U_{t[1:r]} \Sigma_{t[1:r]}^{1 / 2} U_{t[1:r]}^{T}$ with the eigenvalues and eigenvectors obtained in step 2 and step 3.
	\label{alg:VQCMSR}
\end{algorithm}

In step 1, we compute the source domain covariance matrix $H_{s} = \rho_{C_{s}}$ and the target domain covariance matrix $H_{t} = \rho_{C_{t}}$. Then, the Hamiltonian $H_{\tilde{t}} = \eta I - H_{t}$ is determined with a specified constant $\eta$.

In step 2, the ansatz states $| \psi( \lambda_{k}^{(s)} ) \rangle$ are prepared by a quantum circuit with a set of parameters $\{ \theta^{(s)} \}$. Subsequently, the cost function $F_{s}(\lambda_{k}^{(s)})$ is minimized to obtain the optimal ansatz states where the expectation value term $E_{k}^{(s)} = \langle \psi(\lambda_{k}^{(s)}) | H_{s} | \psi(\lambda_{k}^{(s)}) \rangle$, the overlap term $O_{ki}^{(s)} = | \langle \psi(\lambda_{k}^{(s)}) | \psi(\lambda_{i}^{(s)}) \rangle |^{2}$ with the weight coefficient $\alpha_{i}^{(s)}$ for $i = 1, \cdots, k-1$. In the first iteration, we minimize the $F_{s}(\lambda_{1}^{(s)})$ to obtain the ground state $| \psi(\lambda_{1}^{(s)})\rangle$ of $H_{s}$ with the corresponding eigenvalue $\lambda_{1} = E_{1}$. In the second iteration, substitute $| \psi(\lambda_{1}^{(s)}) \rangle$ to $F_{s}(\lambda_{2}^{(s)})$ and minimize it to obtain $| \psi(\lambda_{2}^{(s)}) \rangle$. Then, the iteration continues until $| \psi(\lambda_{D}^{(s)}) \rangle$ is computed by substituting $| \psi(\lambda_{D-1}^{(s)}) \rangle$ to the cost function $F_{s}(\lambda_{D}^{(s)})$. Therefore, $H_{s}$'s eigenstates $| \psi(\lambda_{k}^{(s)}) \rangle$ for $k = 1, \cdots, D$ corresponding to the $D$ eigenvalues can be obtained in $O({1} / {\epsilon^{2}})$~\cite{higgott2019variational}. 

In step 3, the $r$ largest eigenvalues of $H_{t}$ can be obtained similarly by minimizing the cost function $F_{t}(\lambda_{k}^{(t)})$ as exactly the same procedure as in step 2 in time $O({1} / {\epsilon^{2}})$~\cite{higgott2019variational} where the expectation value term $E_{k}^{(t)} = \langle \psi(\lambda_{k}^{(t)}) | \tilde{H}_{t} | \psi(\lambda_{k}^{(t)}) \rangle$, the overlap term $O_{ki}^{(t)} = | \langle \psi(\lambda_{k}^{(t)}) | \psi(\lambda_{i}^{(t)}) \rangle |^{2}$ with the weight coefficient $\alpha_{i}^{(t)}$ for $i = 1, \cdots, k-1$. 

In step 4, the matrices $C_{s}^{1 / 2} = U_{s} \Sigma_{s}^{1 / 2} U_{s}^{T}$ and $C_{t}^{1 / 2} = U_{t[1:r]} \Sigma_{t[1:r]}^{1 / 2} U_{t[1:r]}^{T}$ can be computed by the results of step 2 and step 3. Specifically, the $D$ eigenvalues of $H_{s}$ are the diagonal elements of $\Sigma_{s}$ and the columns of $U_{s}$ are the corresponding $D$ eigenvectors. The diagonal elements of $\Sigma_{t}$ are the $r$ largest eigenvalues of $H_{t}$ and the columns of $U_{t}$ are made up of the corresponding eigenvectors.

(2) The procedure of data decorrelation and alignment can be achieved as Eq.~\eqref{eq:decorrelated X_s} and Eq.~\eqref{eq:aligned X_s} which are actually a variational process of matrix multiplication. In the spirit of Ref.~\cite{bravo2019variational}, we design a matrix-multiplication-based VQCORAL as in Algorithm~\ref{alg:VMM}.

\begin{algorithm}[htb]
	\caption{Matrix-multiplication-based VQCORAL}
	%\LinesNumbered
	\KwIn{Source domain data $X_{s}$ with labels $L_{s}$; target domain data $X_{t}$; $C_{s}^{1/ 2}$ and $C_{t}^{1 / 2}$.}
	\KwOut{Target domain labels $L_{t}$.}
	\emph{step 1}: Prepare the ansatz states $| x_{i}^{(\tilde{s})}(\theta^{(d)}) \rangle$ by parameterized quantum circuits and a set of parameters $\{ \theta^{(d)} \}$ to represent the data point of the decorrelated source domain data $X_{\tilde{s}}$. \\
	\emph{step 2}: Minimize the cost function
	\begin{equation}
		L_{m1} = 1 - \frac{1}{n_{s}} \sum_{i=1}^{n_{s}} \left \vert \frac{\langle x_{i}^{(s)} | C_{s}^{\frac{1}{2}} | x_{i}^{(\tilde{s})}(\theta^{(d)}) \rangle}{\sqrt{\langle x_{i}^{(\tilde{s})} (\theta^{(d)}) | C_{s}^{\frac{1}{2} \dagger} C_{s}^{\frac{1}{2}} | x_{i}^{(\tilde{s})} (\theta^{(d)}) \rangle}} \right \vert^{2} \notag
	\end{equation}
	to obtain the optimal decorrelated source domain data state $| x_{i \ast}^{(\tilde{s})}(\theta^{(d)}) \rangle$. \\ 
	\emph{step 3}: Prepare the ansatz states $| x_{i}^{(\hat{s})}(\theta^{(a)}) \rangle$ by parameterized quantum circuits and a set of parameters $\{ \theta^{(a)} \}$ to represent the data point of the aligned source domain data $X_{\hat{s}}$. \\
	\emph{step 4}: Minimize the cost function
	\begin{equation}
		L_{m2} = 1 - \frac{1}{n_{s}} \sum_{i=1}^{n_{s}} \left \vert \frac{\langle x_{i}^{(\hat{s})}(\theta^{(a)}) | C_{t}^{\frac{1}{2}} | x_{i \ast}^{(\tilde{s})}(\theta^{(d)}) \rangle}{\sqrt{\langle x_{i}^{(\hat{s})} (\theta^{(a)}) | C_{s}^{\frac{1}{2}} C_{s}^{\frac{1}{2} \dagger} | x_{i}^{(\hat{s})} (\theta^{(a)}) \rangle}} \right \vert^{2} \notag 
	\end{equation}
	to obtain the optimal aligned source domain data state $| x_{i \ast}^{(\hat{s})}(\theta^{(a)}) \rangle$. \\
	\emph{step 5}: Invoke a classifier to predict the target labels $L_{t} = Classifier(X_{\hat{s}}, L_{s}, X_{t})$.
	\label{alg:VMM}
\end{algorithm}

In step 1, the quantum ansatz states $| x_{i}^{(\tilde{s})}(\theta^{(d)}) \rangle$ representing the decorrelated source domain data point $x_{i}^{(\tilde{s})}$ are designed by parameterized quantum circuits with a set of parameters $\{ \theta^{(d)} \}$.

In step 2, the state 
\begin{equation}
	| \psi_{1} \rangle = \frac{C_{s}^{\frac{1}{2}} | x_{i}^{(\tilde{s})} (\theta^{(d)}) \rangle}{ \sqrt{\langle x_{i}^{(\tilde{s})} (\theta^{(d)}) | C_{s}^{\frac{1}{2} \dagger} C_{s}^{\frac{1}{2}} | x_{i}^{(\tilde{s})} (\theta^{(d)}) \rangle}}
	\label{eq:psi_1}
\end{equation}
is defined to be proportional to $| x_{i}^{(s)} \rangle$ with a set of parameters $\{ \theta^{(d)} \}$. Thus, the optimal quantum ansatz states $| x_{i \ast}^{(\tilde{s})}(\theta^{(d)}) \rangle$ representing the decorrelated source domain data can be obtained by minimizing the cost function
\begin{equation}
	L_{m1} = 1 - \frac{1}{n_{s}} \sum_{i=1}^{n_{s}} \vert \langle x_{i}^{(s)} | \psi_{1} \rangle \vert^{2}
	\label{eq:L_m1} 
\end{equation}  
in time $O(\kappa_{s} / \epsilon)$~\cite{bravo2019variational} where $\kappa_{s}$ is the conditional number of $C_{s}^{\frac{1}{2}}$.

In step 3, the ansatz states $| x_{i}^{(\hat{s})}(\theta^{(a)}) \rangle$ are prepared by parameterized quantum circuits with a set of parameters $\{ \theta^{(a)} \}$.

In step 4, define the state
\begin{equation}
	| \psi_{2} \rangle = \frac{C_{t}^{\frac{1}{2} \dagger} | x_{i}^{(\hat{s})} (\theta^{(a)}) \rangle}{ \sqrt{\langle x_{i}^{(\hat{s})} (\theta^{(a)}) | C_{s}^{\frac{1}{2}} C_{s}^{\frac{1}{2} \dagger} | x_{i}^{(\hat{s})} (\theta^{(a)}) \rangle}}
	\label{eq:psi_2}
\end{equation}
with a set of parameters $\{ \theta^{(a)} \}$.
Then, we align the decorrelated source domain data $X_{\hat{s}}$ by minimizing 
\begin{equation}
	L_{m2} = 1 - \frac{1}{n_{s}} \sum_{i=1}^{n_{s}} \vert \langle \psi_{2} | x_{i \ast}^{(\tilde{s})}(\theta^{(d)}) \rangle \vert^{2}
\end{equation}
in time $O(\kappa_{t} / \epsilon)$~\cite{bravo2019variational} where $\kappa_{t}$ is the conditional number of $C_{t}^{-\frac{1}{2}}$. The data alignment procedure is actually aims to generate the state $| \psi_{2} \rangle$ to be proportional to $| x_{i}^{(\hat{s})}(\theta^{(a)}) \rangle$.

\begin{figure*}
	\centering
	\includegraphics[width=\textwidth]{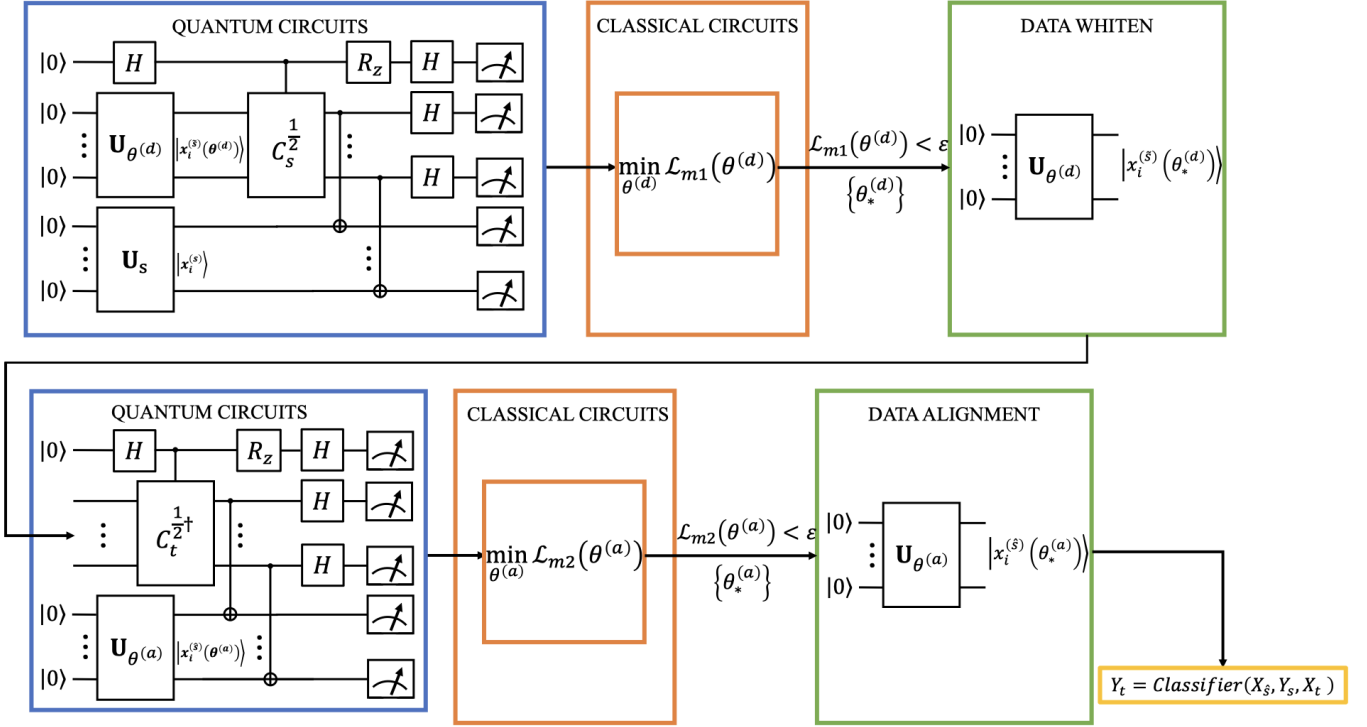}
	\caption{The schematic diagram of the matrix-multiplication-based VQCORAL}
	\label{fig:MM-based VQCORAL}
\end{figure*}

In step 5, the classifier such as the local classifier, the nearest neighbor algorithm, or the global classifier, the support vector machine, will be applied to $X_{\hat{s}}$ with $L_{s}$ and $X_{t}$ to predict the target labels $L_{t}$. The whole procedure of the matrix-multiplication-based VQCORAL is as presented in Fig.~\ref{fig:MM-based VQCORAL}.

\section{Numerical experiments}
\label{sec:numerical experiments}
In this section, three numerical experiments are presented to demonstrate the feasibility and efficiency of the VQCORAL. The no adaptation model (NA), the classical CORAL, the VQCORAL are applied to the synthetic data sets, the synthetic-Iris data sets and the handwritten digit data sets respectively to evaluate their performance. According to the simulation results, the VQCORAL can achieve comparable or even better performance than the classical CORAL. The VQCORAL is simulated on a classical computer using the Python programming language and the Scikit-learn machine learning library~\cite{scikit-learn}. The code and the selected parameters can be found in Ref.~\cite{code}. 

\subsection{Basic settings}
\begin{figure}[b]
	\centering
	\includegraphics[width=\columnwidth]{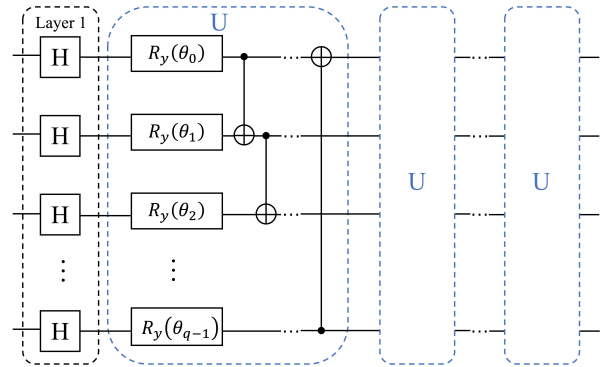}	
	\caption{The variational quantum circuit for preparing $\textbf{U}_{\theta}$ where $q = \log D$}
	\label{fig:U_theta}
\end{figure}

The no adaptation model (NA) is set as the baseline model. In addition, the classical CORAL is also selected as a performance comparison of the VQCORAL. As to the VQCORAL, we design parameterized quantum circuits with hierarchical structures. Specifically, we apply the Hadamard operation on each register respectively as the first layer. Then, we alternately apply the rotation layer constructed by the $R_{y}$ gate on each qubit and the entanglement layer constructed by the controlled-not gate on each two qubits to introduce the parameters and entanglement as shown in Fig.~\ref{fig:U_theta}. The classical optimization algorithm, the AdaGrad~\cite{duchi2011adaptive}, is selected to optimize the cost function. 

\subsection{Synthetic data}
\label{subsec:experiment 1}
\begin{figure*}
	\centering
	\includegraphics[width=0.9\columnwidth]{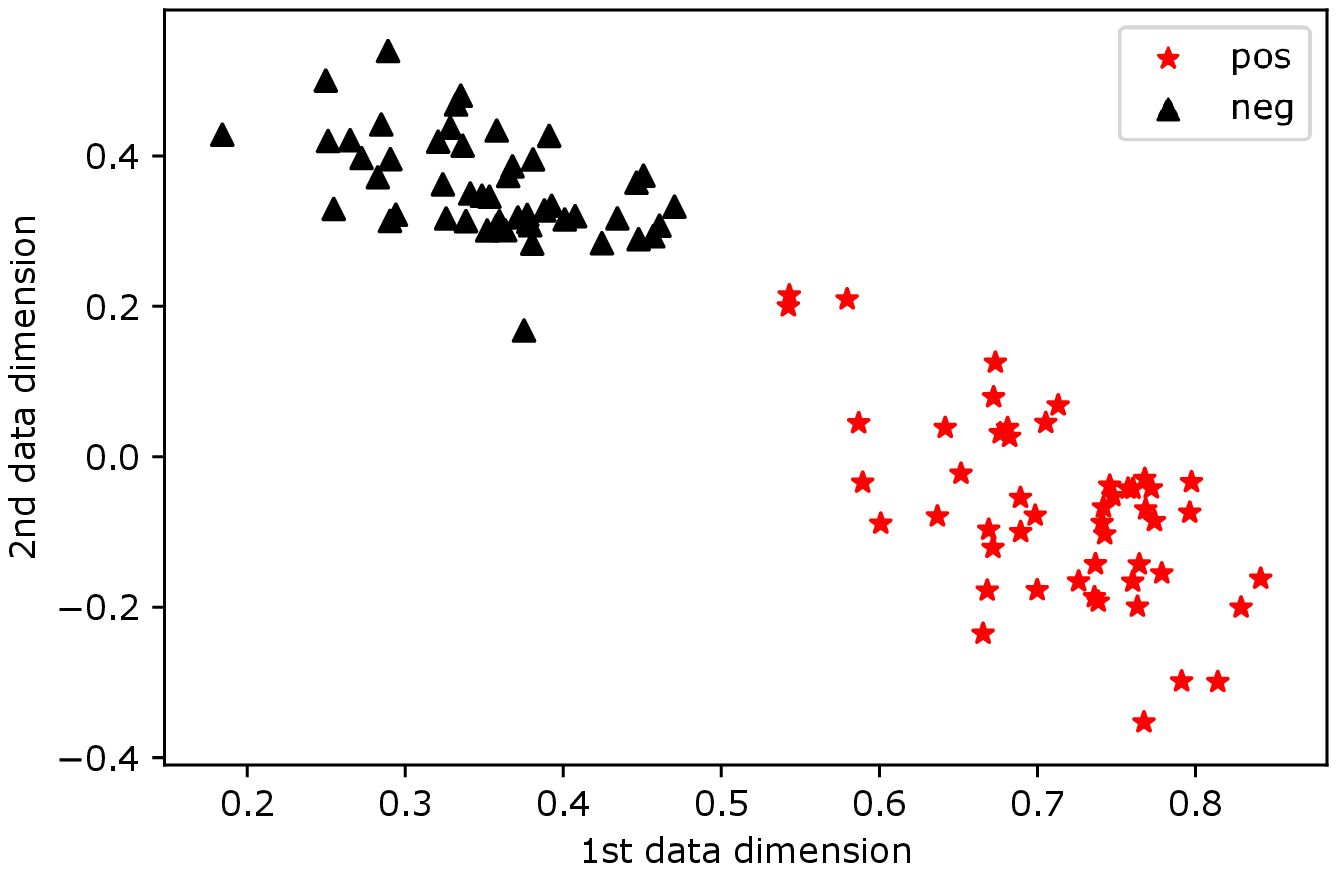}
	\quad
	\includegraphics[width=0.9\columnwidth]{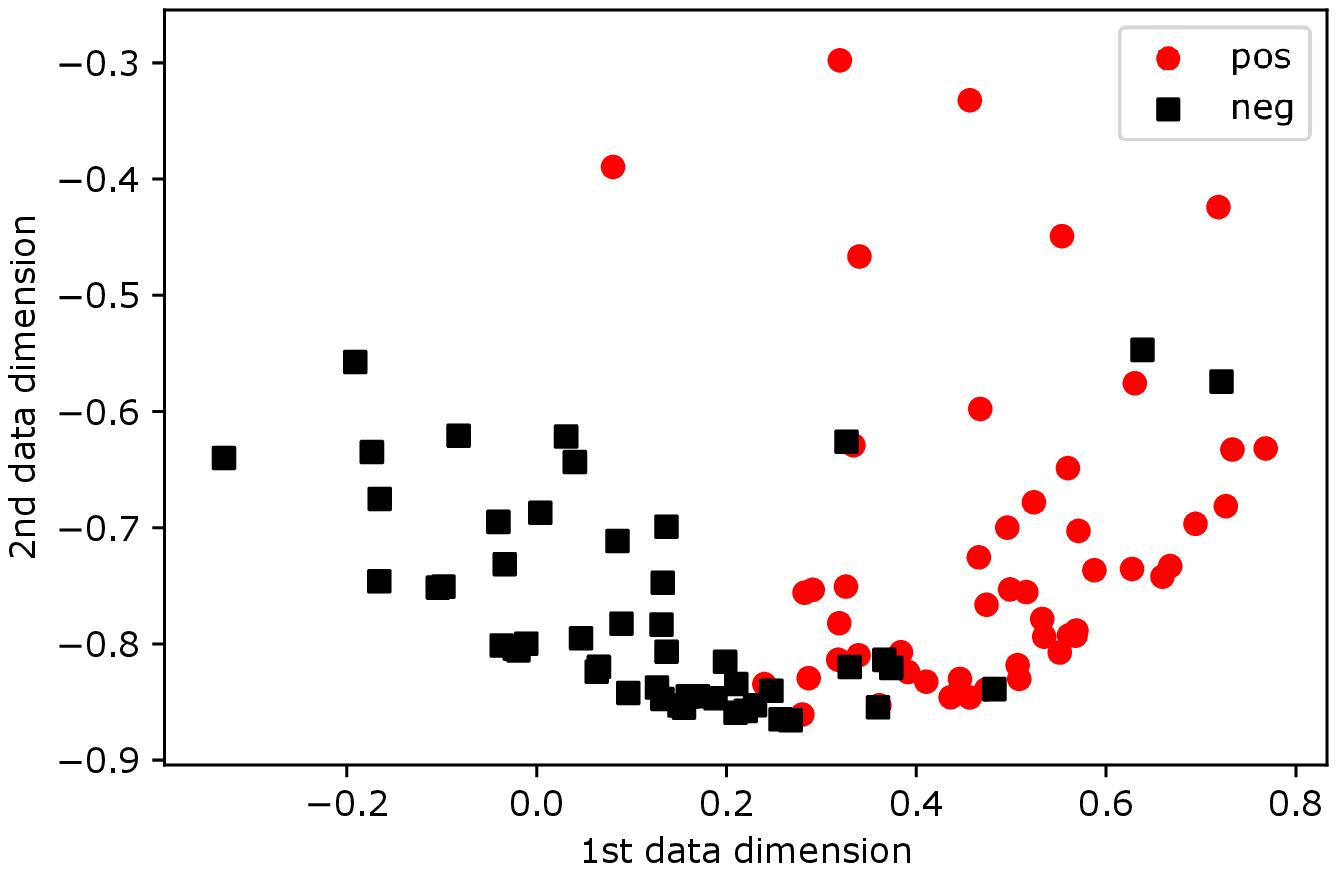} \\
	{\bf (a)} \hspace{8 cm} {\bf (b)} 
	
	\caption{The visualization of the $D_{1}$ dataset and the $D_{2}$ dataset respectively. (a) $D_{1}$ dataset; (b) $D_{2}$ dataset.}
	\label{fig:D1_D2}
\end{figure*}

In the first numerical experiment, we select two synthetic data sets $D_{1} \sim \mathcal{N}(\mu_{1}^{(1)}=\mu_{2}^{(1)}=0, \sigma_{1}^{(1)}=\sigma_{2}^{(1)}=1)$ and $D_{2} \sim \mathcal{N}(\mu_{1}^{(2)}=\mu_{2}^{(2)}=0, \sigma_{1}^{(2)}=\sigma_{2}^{(2)}=2)$ depicted in Fig.~\ref{fig:D1_D2} as the source and target domain data sets alternately. Both $X_{s}$ and $X_{t}$ contain 100 four-dimensional data points distributed in two different classes.

\begin{figure*}
	\centering
	\includegraphics[width=0.65\columnwidth]{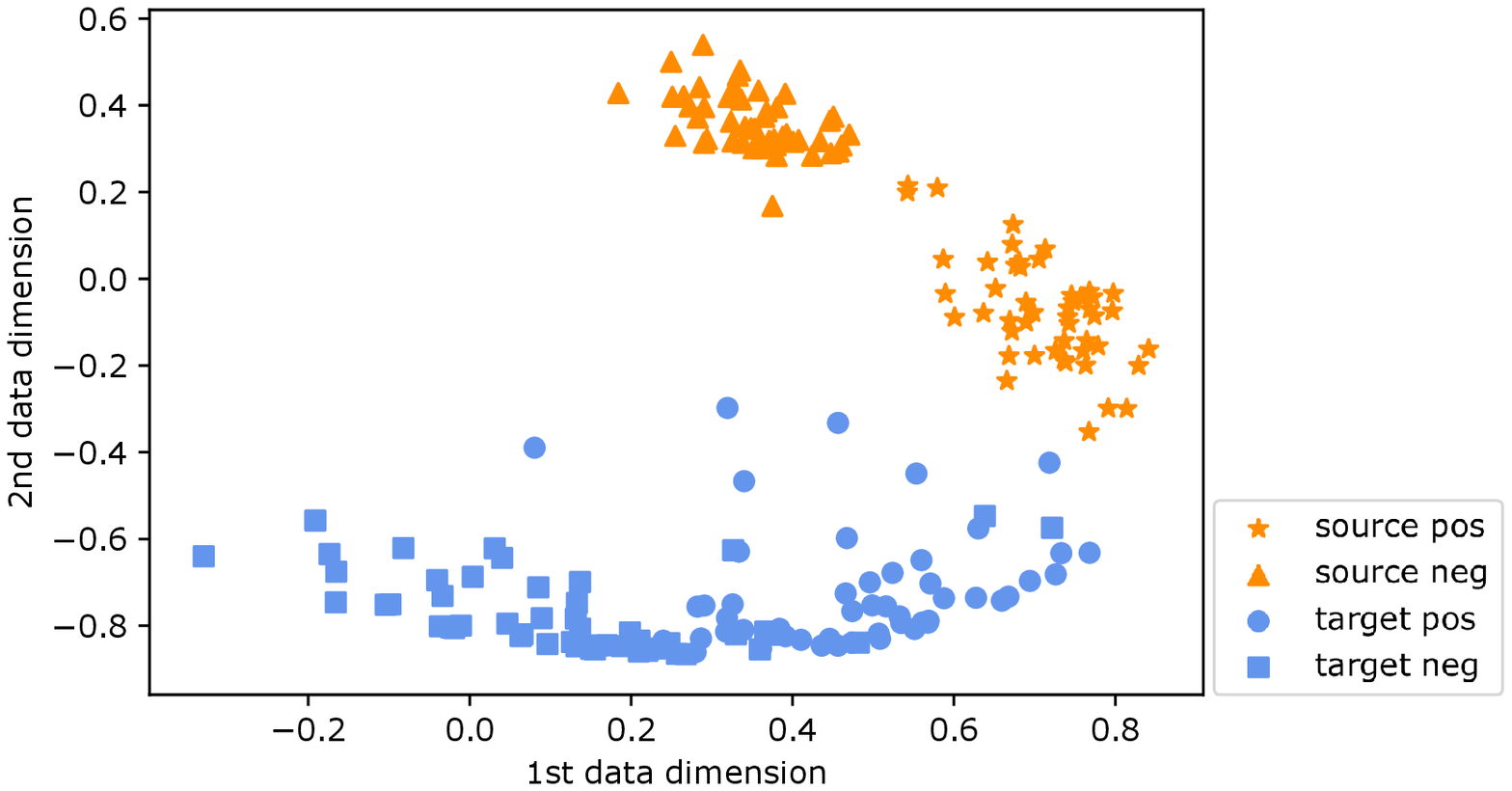}
	\quad
	\includegraphics[width=0.65\columnwidth]{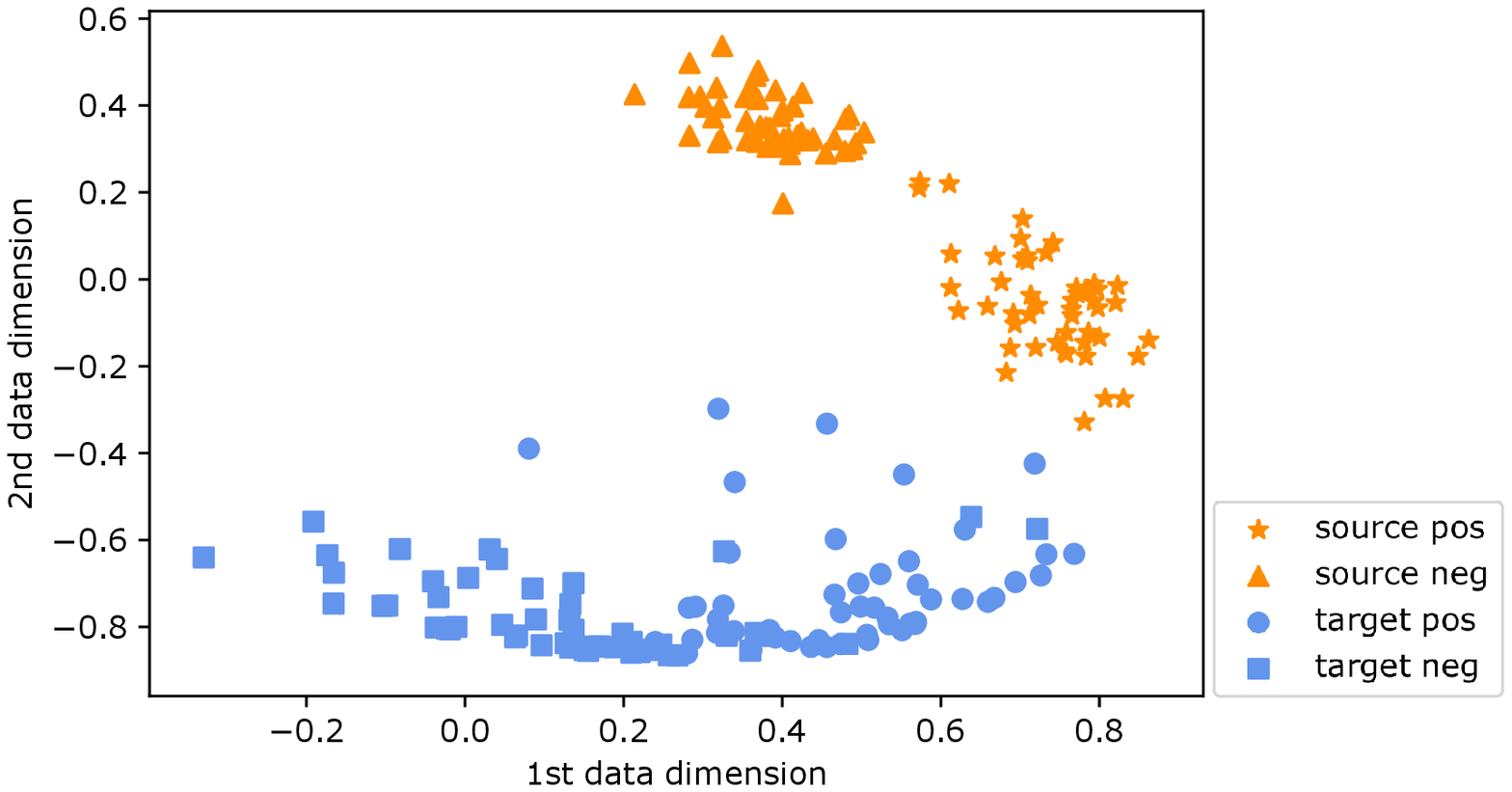}
	\quad
	\includegraphics[width=0.65\columnwidth]{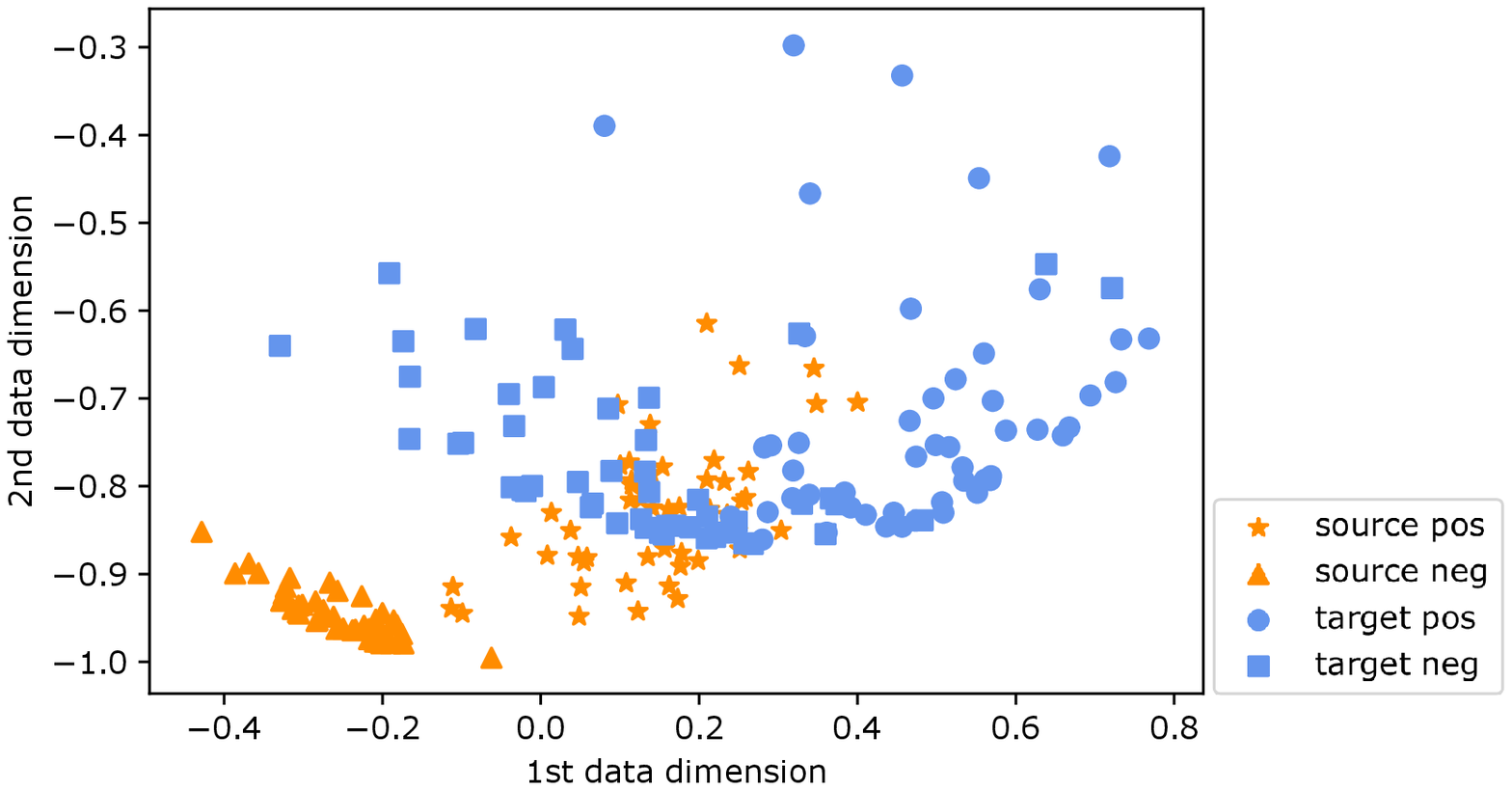}\\
	{\bf (a)} \hspace{5.5 cm} {\bf (b)} \hspace{5.5 cm} {\bf (c)}\\
	\includegraphics[width=0.65\columnwidth]{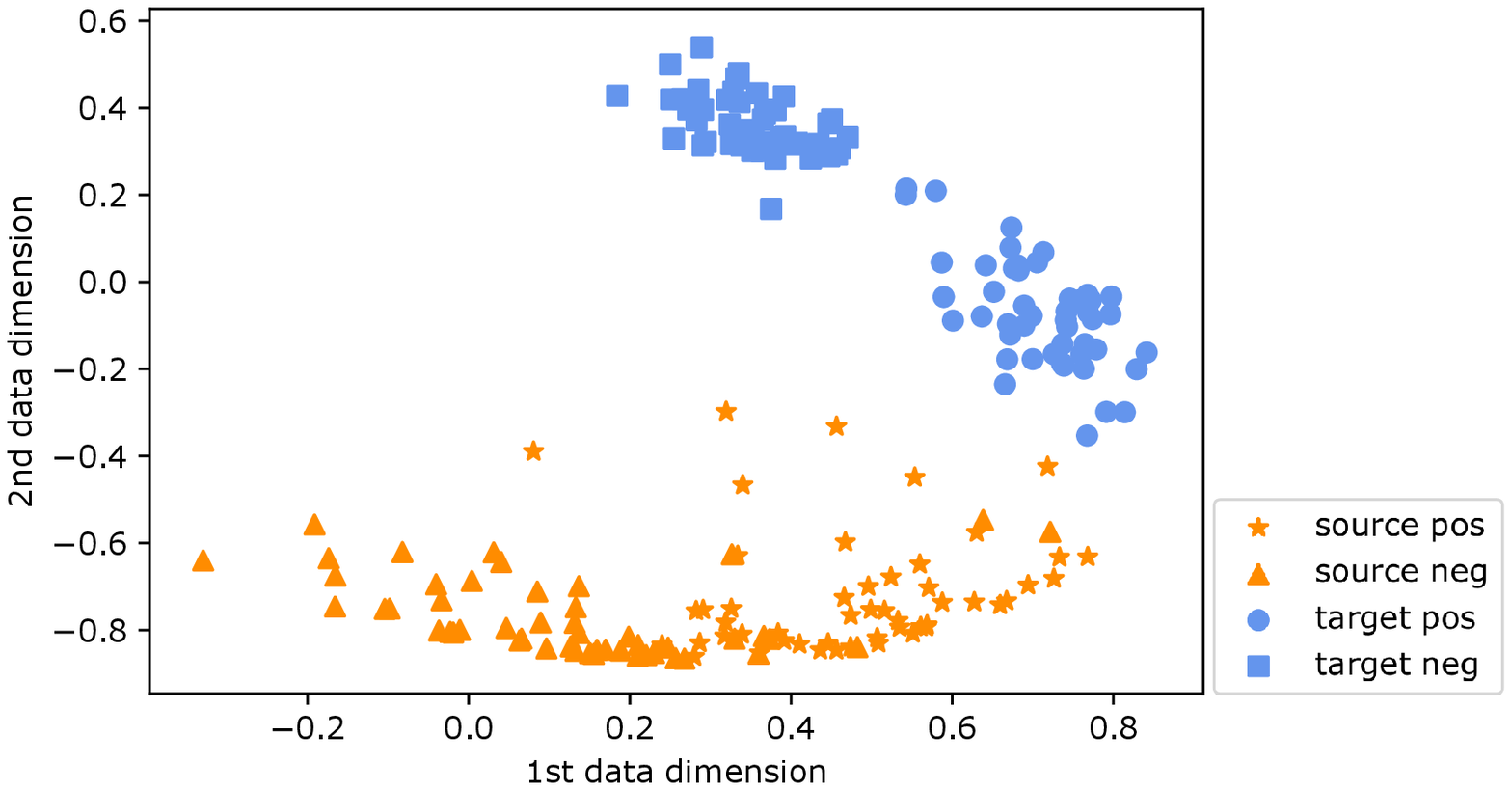}
	\quad
	\includegraphics[width=0.65\columnwidth]{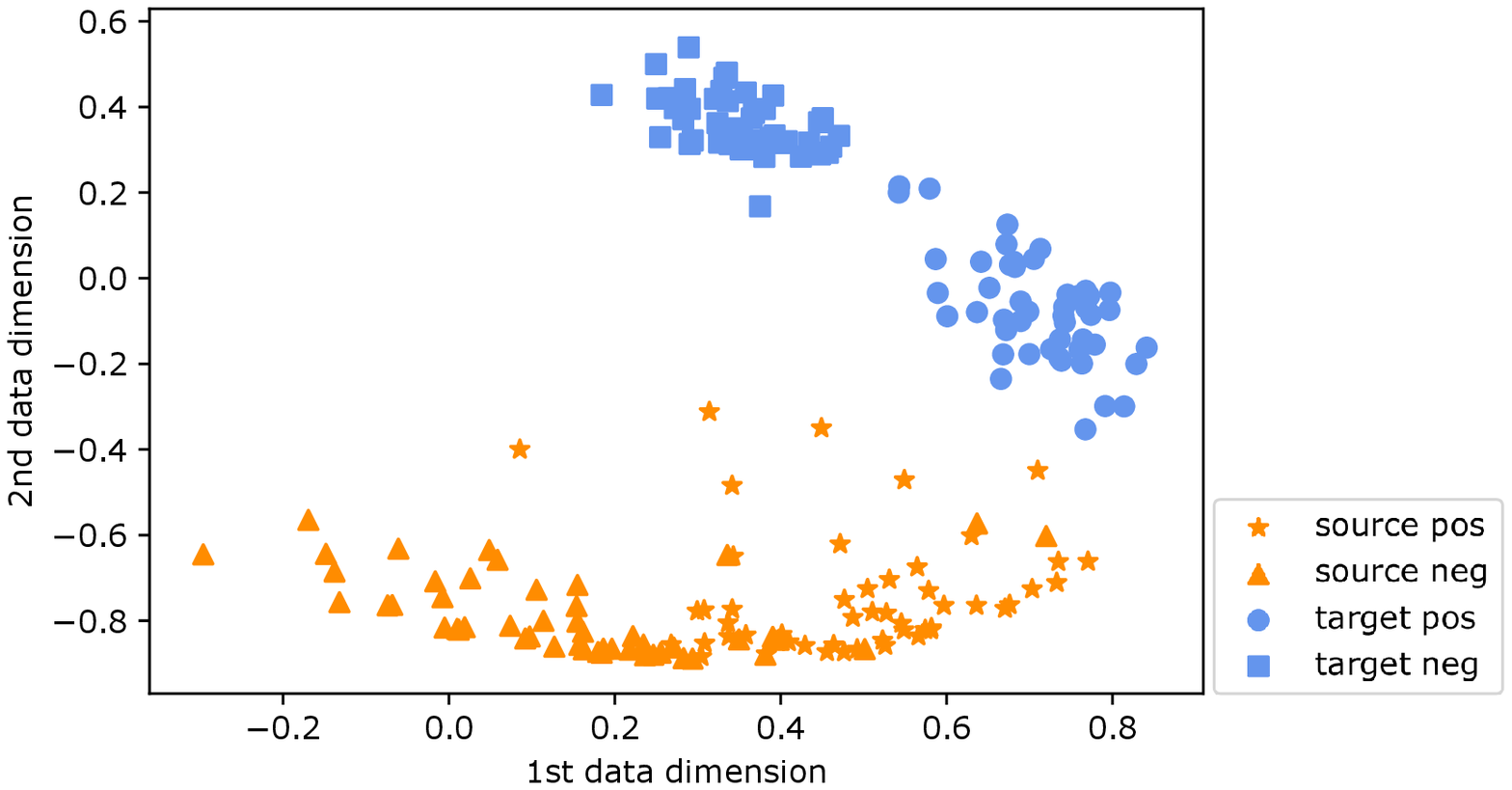}
	\quad
	\includegraphics[width=0.65\columnwidth]{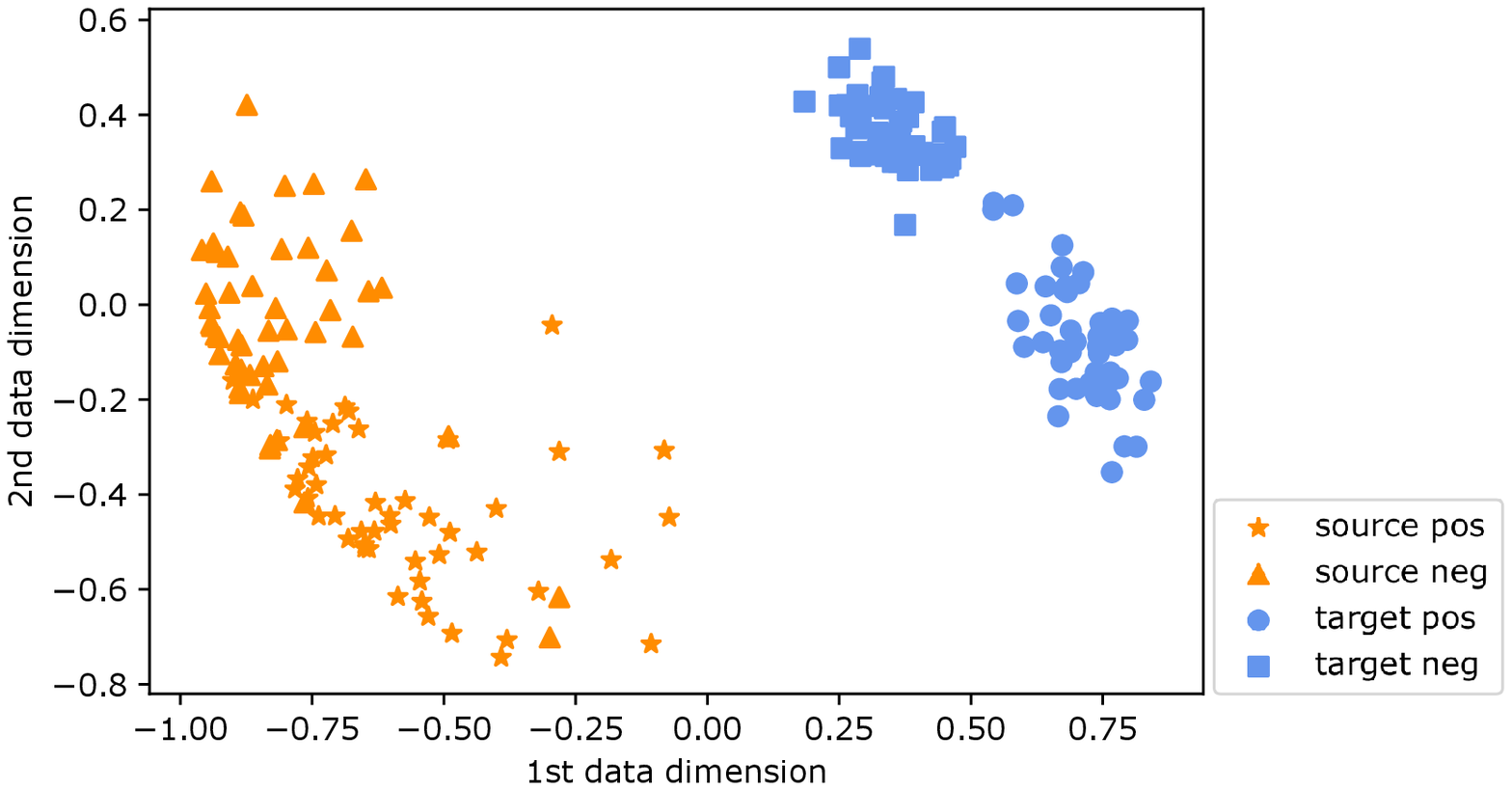}\\
	{\bf (d)} \hspace{5.5 cm} {\bf (e)} \hspace{5.5 cm} {\bf (f)}
	
	\caption{The visualization of the results of experiment. (a) NA model in $D_{1} \rightarrow D_{2}$ dataset; (b) Classical CORAL model in $D_{1} \rightarrow D_{2}$ dataset; (c) VQCORAL model in $D_{1} \rightarrow D_{2}$ dataset; (d) NA model in $D_{2} \rightarrow D_{1}$ dataset; (e) Classical CORAL model in $D_{2} \rightarrow D_{1}$ dataset; (f) VQCORAL model in $D_{2} \rightarrow D_{1}$ dataset;}
	\label{fig:result_1}
\end{figure*}

The design of the VQCORAL in this experiment is a $2$-qubit $8$-layer quantum circuit. The simulation results of the NA, the classical CORAL, and the VQCORAL applied to the $D_{1} \rightarrow D_{2}$ task and the $D_{2} \rightarrow D_{1}$ task are presented in Table.~\ref{tab:D_1_D_2}. In addition, the visualization of the results of this experiment is presented in Fig.~\ref{fig:result_1}

As shown in Table.~\ref{tab:D_1_D_2}, for both the $D_{1} \rightarrow D_{2}$ and the $D_{2} \rightarrow D_{1}$ tasks, it is obvious that the NA (baseline model) can not achieve a relative high accuracy. However, the performance of the classical CORAL is comparable to the NA meaning that the classical CORAL may not play the role of domain adaptation as we expected in some cases. Compared with the classical CORAL and the NA, the VQCORAL model achieves significantly better performance.

\begin{table}[htb]
	\caption{\label{tab:D_1_D_2}%
	Accuracies of the NA, the classical CORAL, and the VQCORAL applied on the synthetic data sets $D_{1}$ and $D_{2}$
	}
	\begin{ruledtabular}
		\begin{tabular}{ccc}
			\textrm{}&
			\textrm{$D_{1} \rightarrow D_{2}$}&
			\textrm{$D_{2} \rightarrow D_{1}$}\\
			\colrule
			NA & 50\% & 50\%\\
			Classical CORAL & 50\% & 50\%\\
			VQCORAL & \textbf{90\%} & \textbf{97\%}\\
		\end{tabular}
	\end{ruledtabular}
\end{table}

\subsection{Synthetic and Iris data}
\label{subsec:data set 2}
In the second experiment, the synthetic data set $D_{3} \sim \mathcal{N} (\mu_{1}^{(3)} = \mu_{2}^{(3)} = \mu_{3}^{(3)} = 0, \sigma_{1}^{(3)} = \sigma_{2}^{(3)} = \sigma_{3}^{(3)} = 1)$ and the Iris data set~\cite{fisher1936use, anderson1936species} depicted in Fig~\ref{D3_Iris} are selected as the source and target domain data sets alternately. Both the $D_{3}$ and the Iris data set contains $150$ samples evenly distributed in three different classes.

\begin{figure*}
	\centering
	\includegraphics[width=0.9\columnwidth]{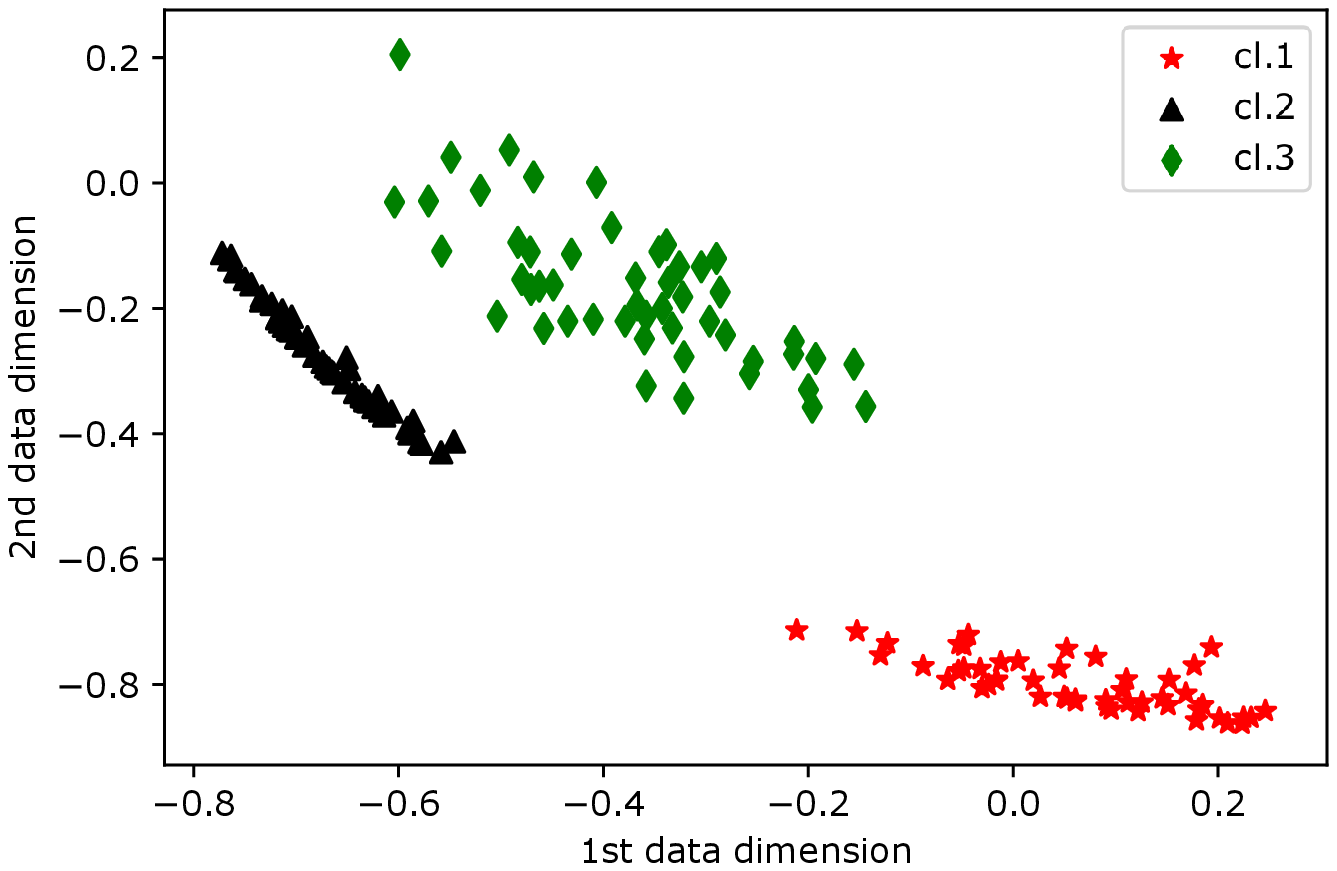}
	\quad
	\includegraphics[width=0.9\columnwidth]{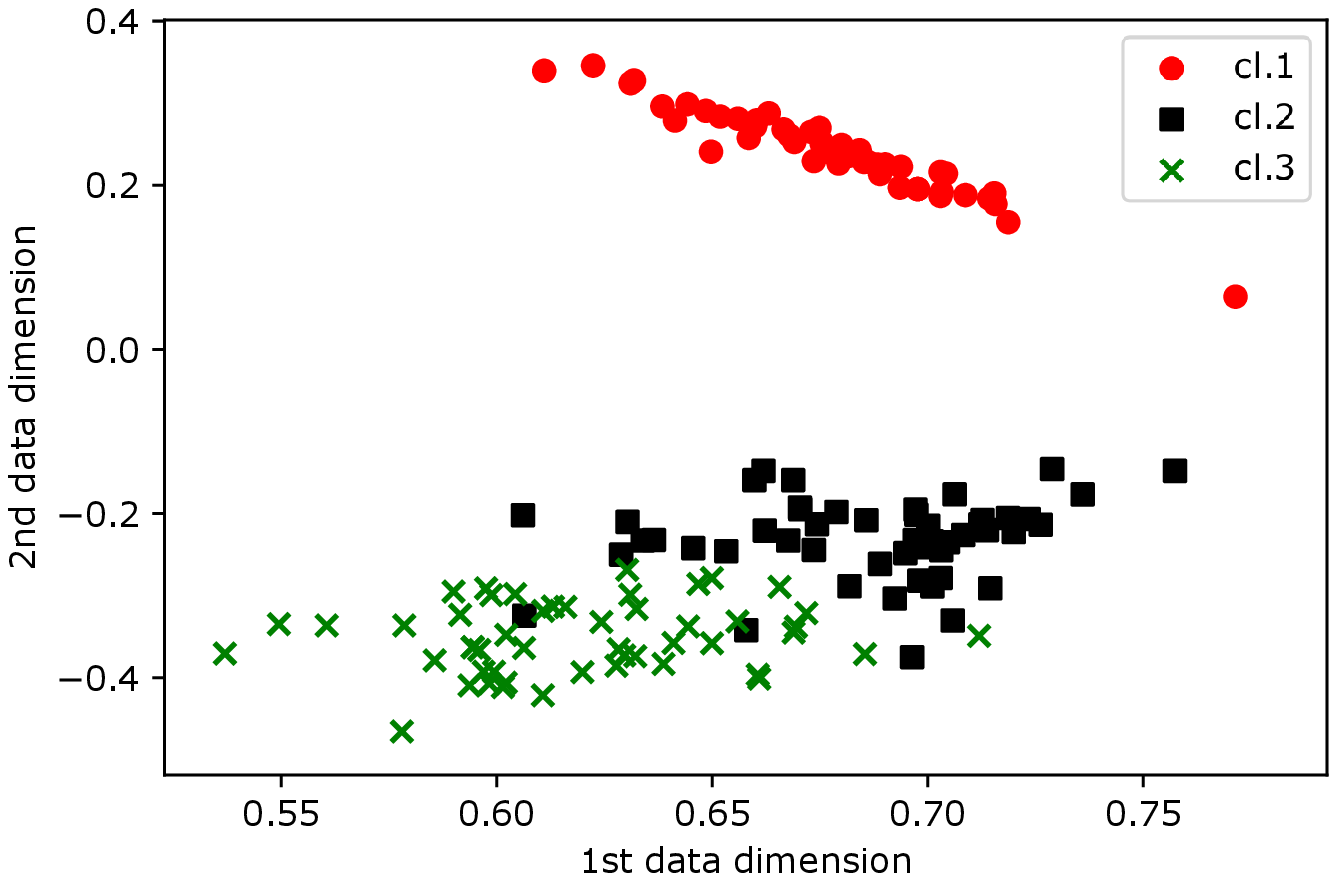} \\
	{\bf (a)} \hspace{8 cm} {\bf (b)} 
	
	\caption{The visualization of the $D_{3}$ dataset and the Iris dataset respectively. (a) $D_{1}$ dataset; (b) $D_{2}$ dataset.}
	\label{fig:D3_Iris}
\end{figure*}

\begin{figure*}
	\centering
	\includegraphics[width=0.65\columnwidth]{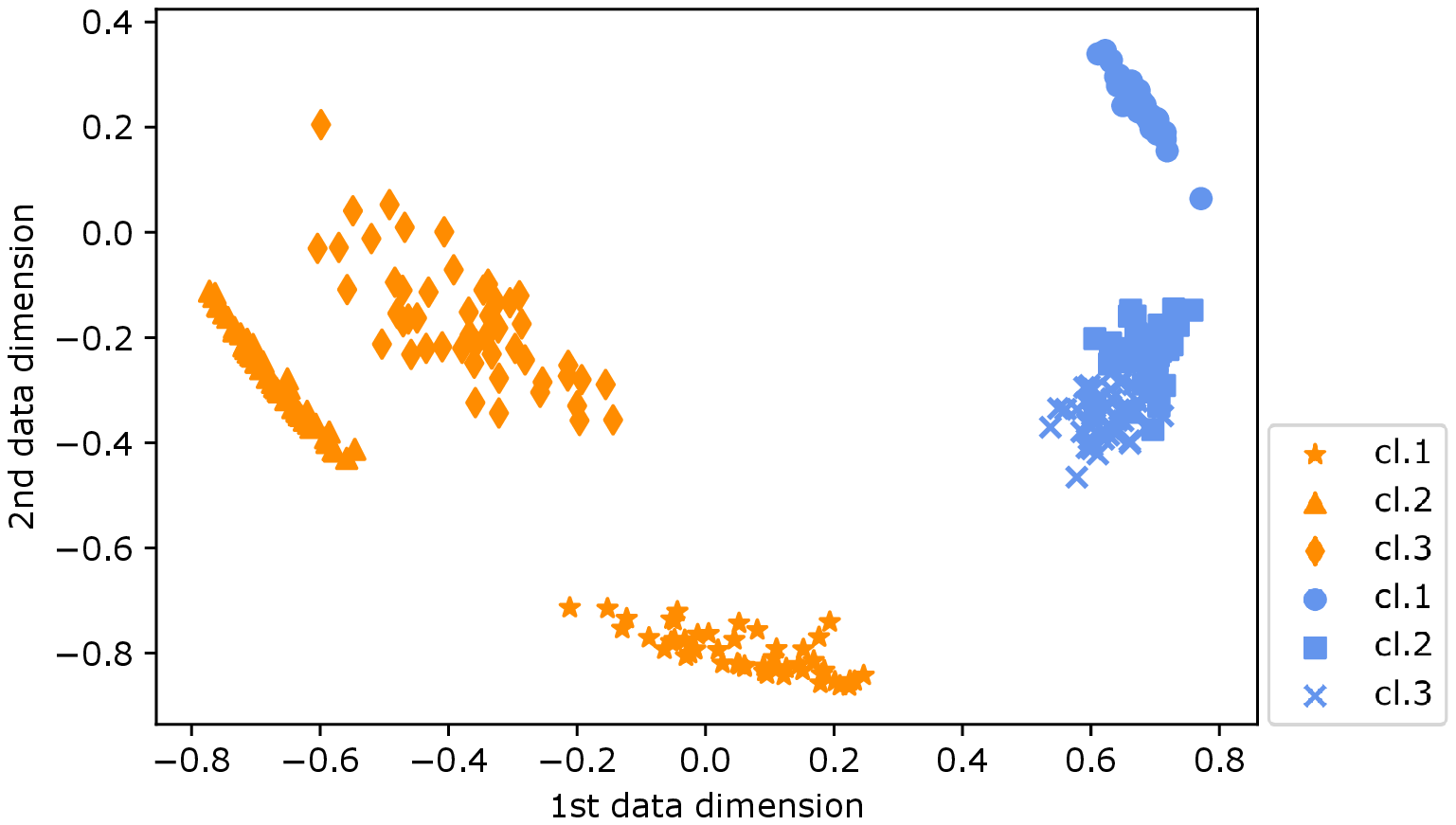}
	\quad
	\includegraphics[width=0.65\columnwidth]{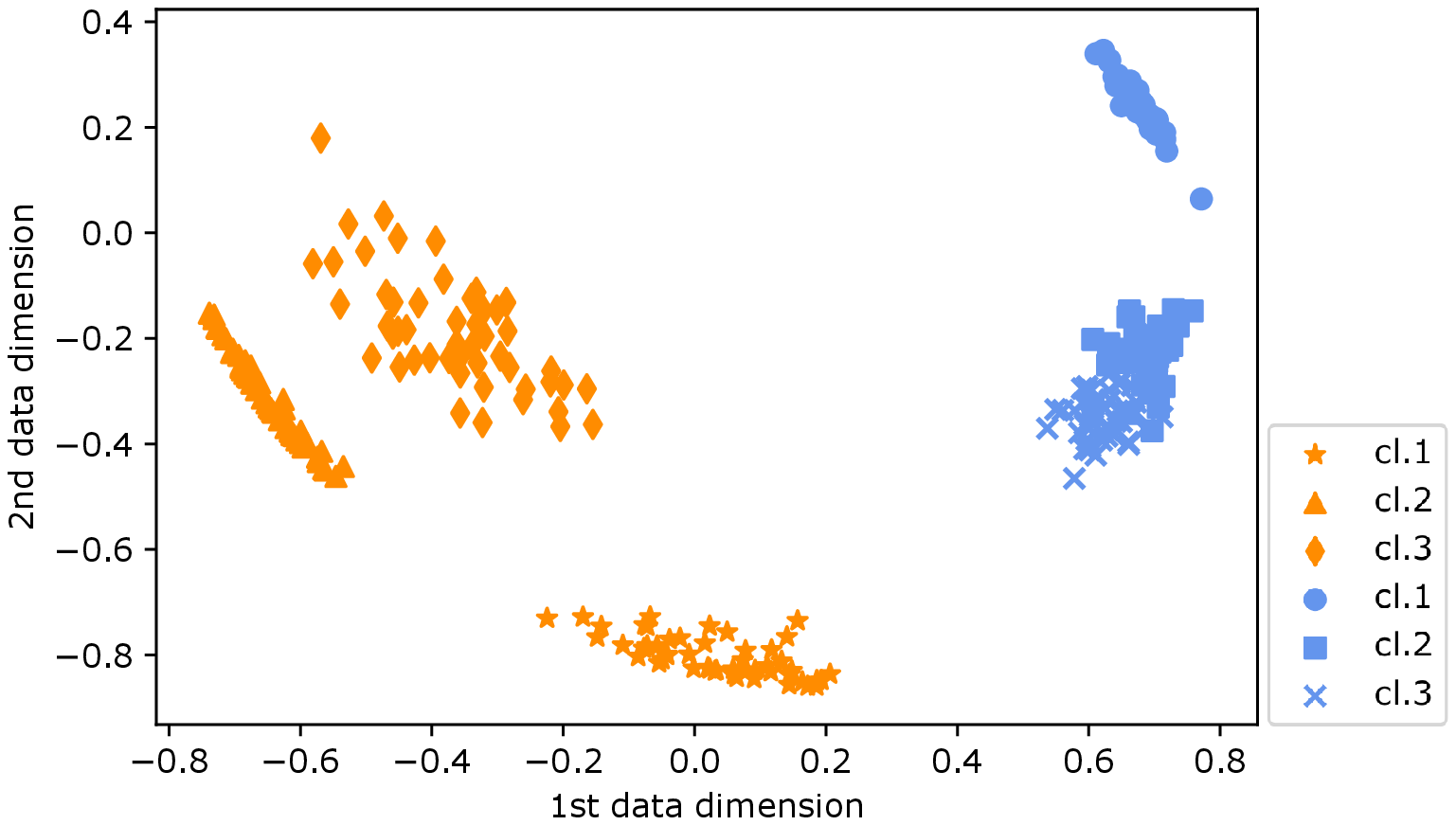}
	\quad
	\includegraphics[width=0.65\columnwidth]{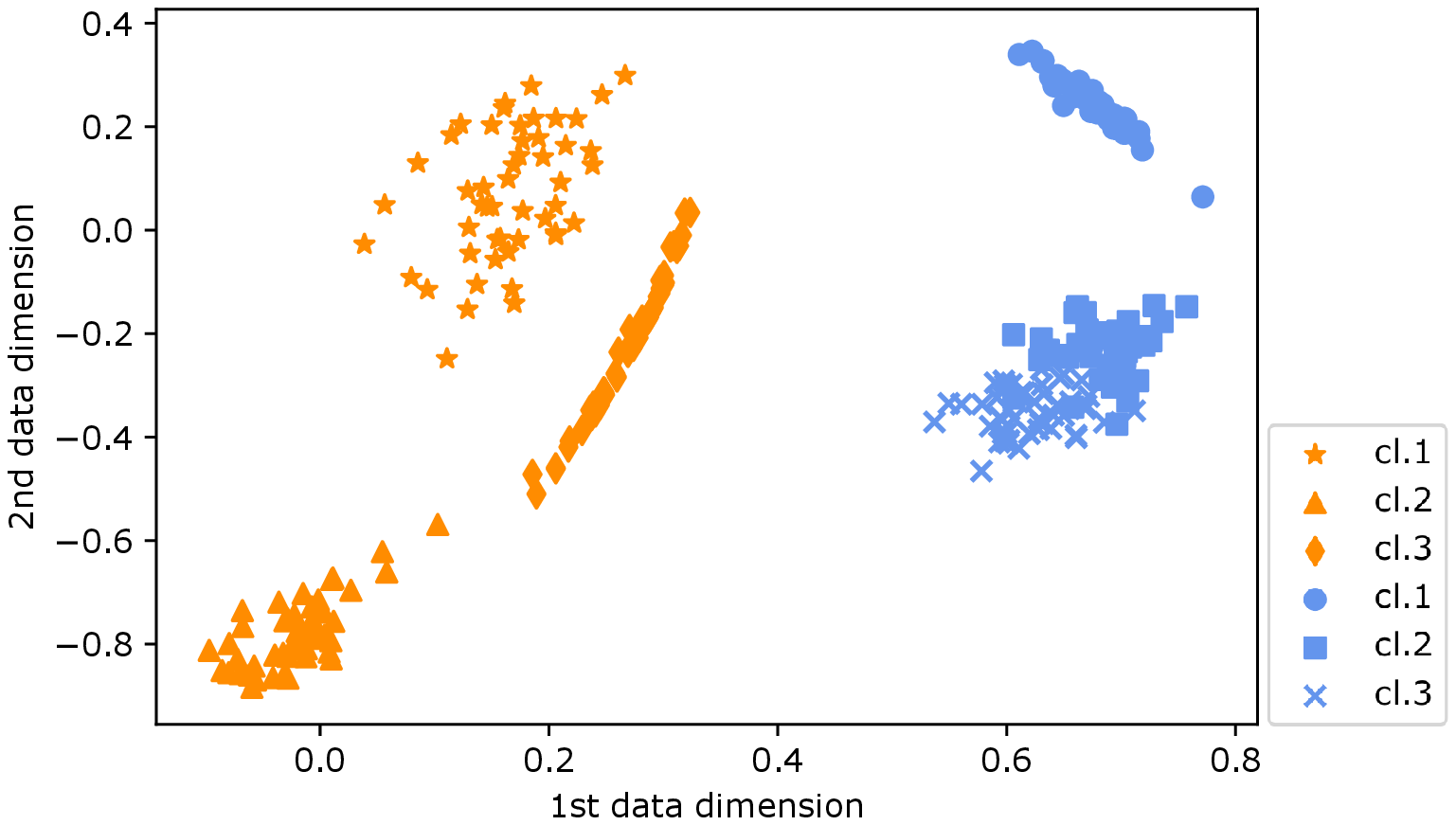}\\
	{\bf (a)} \hspace{5.5 cm} {\bf (b)} \hspace{5.5 cm} {\bf (c)}\\
	\includegraphics[width=0.65\columnwidth]{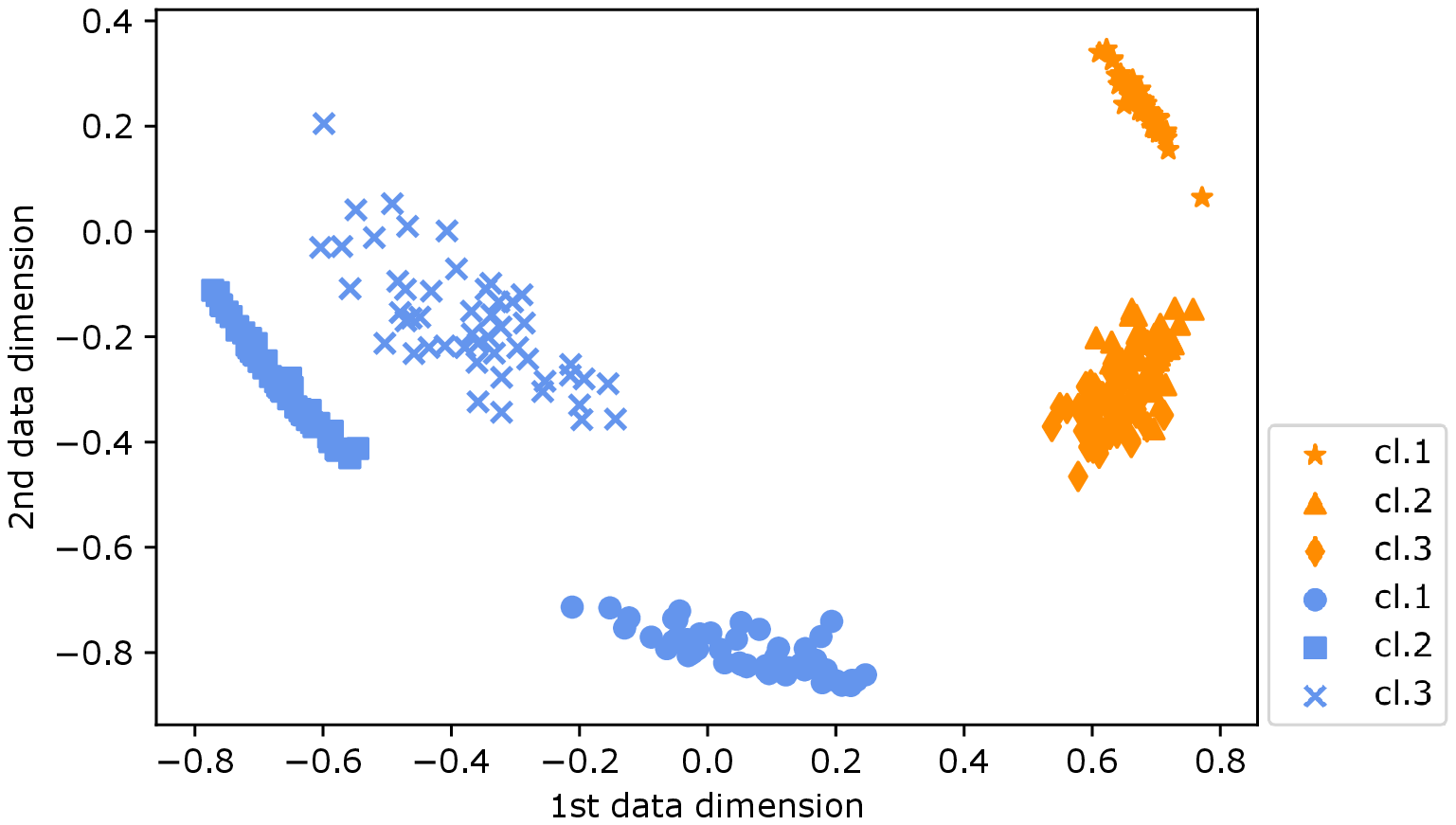}
	\quad
	\includegraphics[width=0.65\columnwidth]{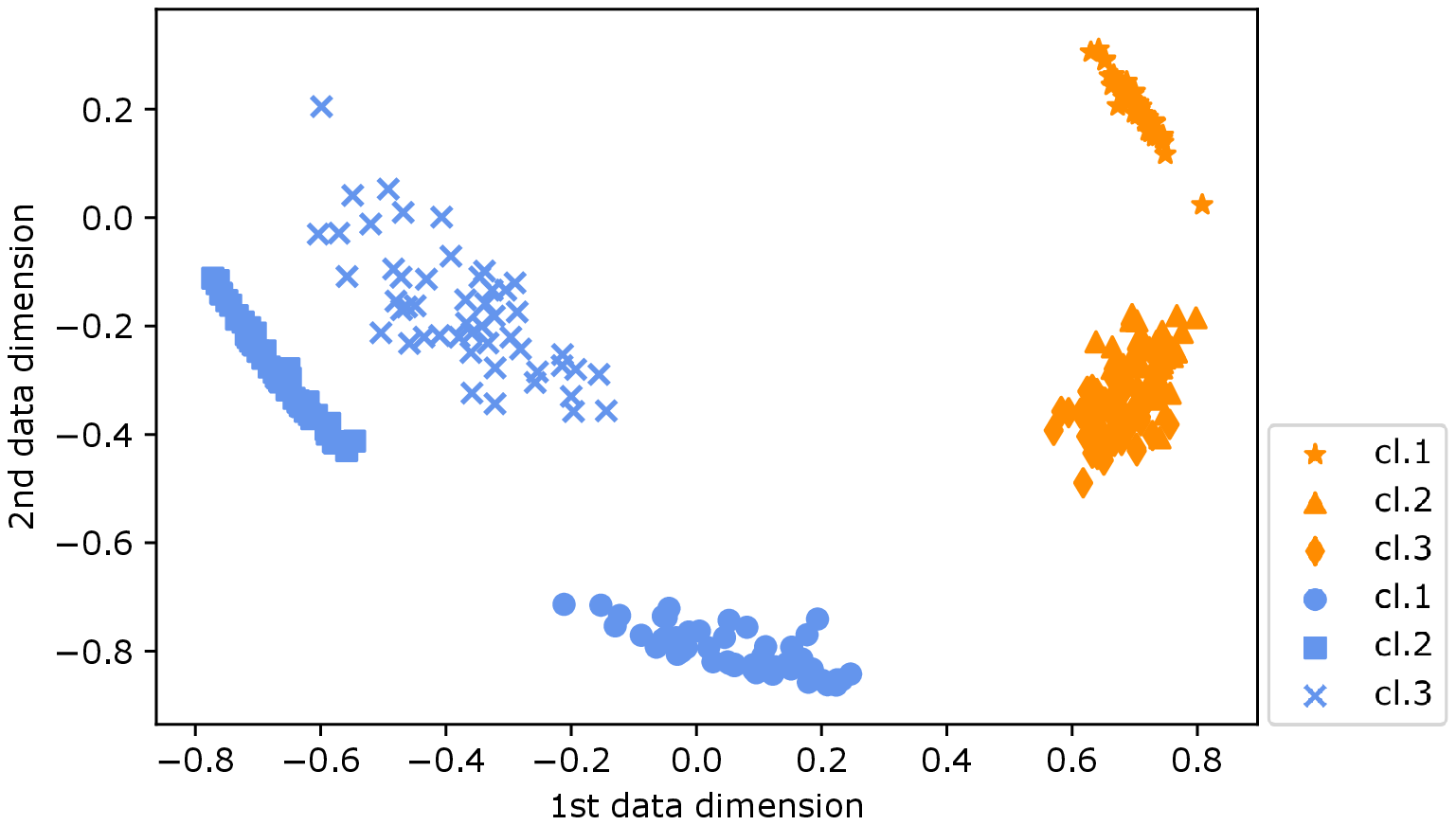}
	\quad
	\includegraphics[width=0.65\columnwidth]{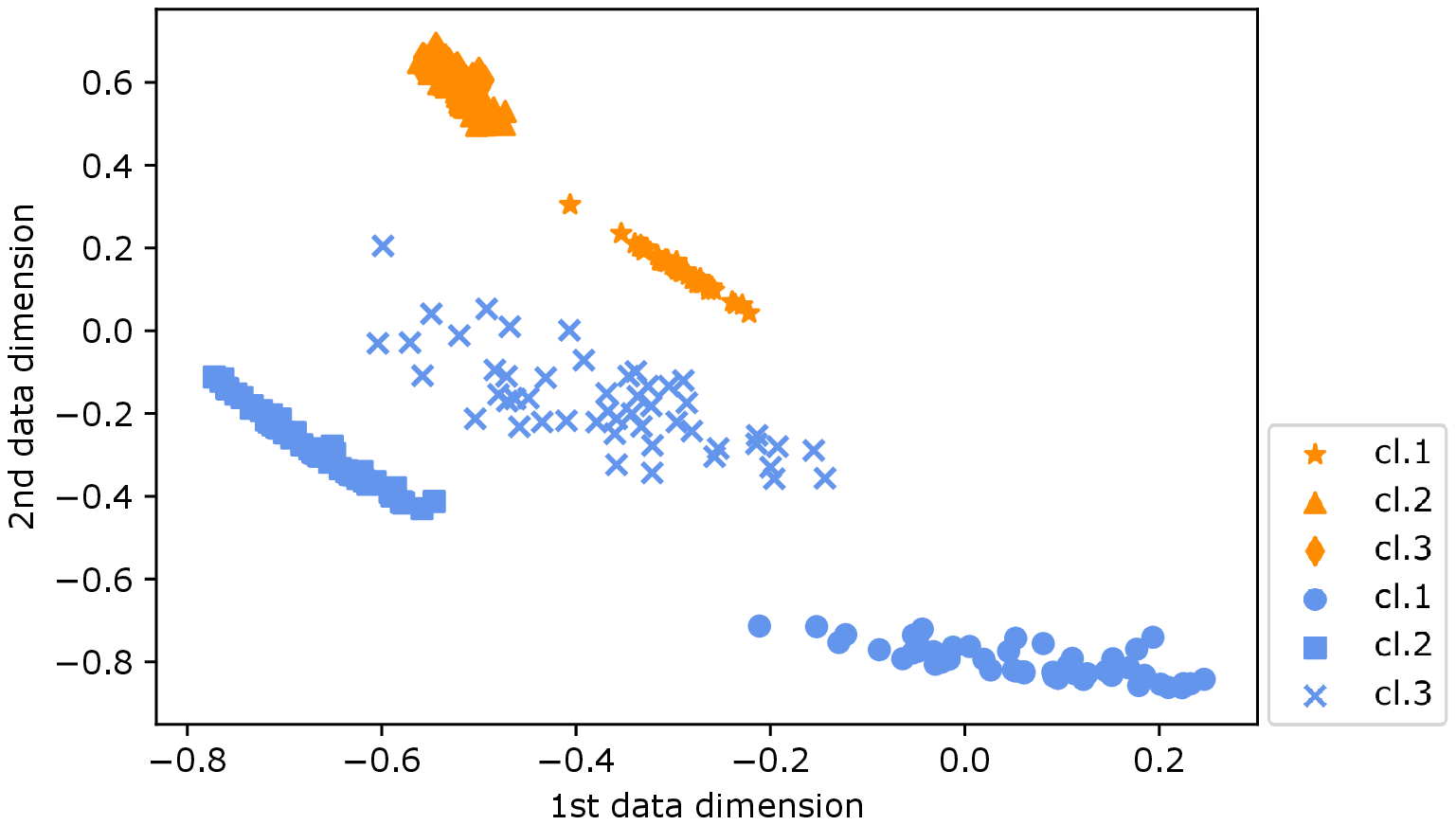}\\
	{\bf (d)} \hspace{5.5 cm} {\bf (e)} \hspace{5.5 cm} {\bf (f)}
	
	\caption{The visualization of the results of experiment. (a) NA model in $D_{3} \rightarrow Iris$ dataset; (b) Classical CORAL model in $D_{3} \rightarrow Iris$ dataset; (c) VQCORAL model in $D_{3} \rightarrow Iris$ dataset; (d) NA model in $Iris \rightarrow D_{3}$ dataset; (e) Classical CORAL model in $Iris \rightarrow D_{3}$ dataset; (f) VQCORAL model in $Iris \rightarrow D_{3}$ dataset;}
	\label{fig:result_2}
\end{figure*}

The model adopted by the VQCORAL in this experiment is a $2$-qubit $8$-layer parameterized quantum circuit. The NA, the classical CORAL, and the VQCORAL are applied to the $D_{3} \rightarrow$ Iris task and the Iris $\rightarrow D_{3}$ task resulting in the results in Table.~\ref{tab:D_3_Iris}. The visualization of the results of this experiments is presented in Fig.~\ref{fig:result_2}.

\begin{table}[htb]
	\caption{\label{tab:D_3_Iris}%
	Accuracies of the NA, the classical CORAL, the VQCORAL applied on the synthetic data set $D_{3}$ and the Iris data set.
	}
	\begin{ruledtabular}
		\begin{tabular}{ccc}
			\textrm{}&
			\textrm{$D_{3} \rightarrow$ Iris}&
			\textrm{Iris $\rightarrow D_{3}$}\\
			\colrule
			NA & 33.3\% & 4\%\\
			Classical CORAL & 33.3\% & 14\%\\
			VQCORAL & \textbf{66.6\%} & \textbf{72.7\%}\\
		\end{tabular}
	\end{ruledtabular}
\end{table}

As in Table.~\ref{tab:D_1_Iris}, the accuracy of both the NA and the classical CORAL is $33.3 \%$ for the $D_{3} \rightarrow$ Iris task which is worse than the $66.6 \%$ accuracy of the VQCORAL. For the Iris $\rightarrow D_{3}$ task, the accuracy of the NA is only $4 \%$. The classical CORAL shows improvement with $14 \%$ accuracy. The VQCORAL achieves significant performance improvement with the accuracy of $72.7 \%$ indicating that the VQCORAL can exhibit more powerful expressivity in some specific tasks. 

\subsection{Handwritten digit data}
\label{subsec:data set 3}
The MNIST~\cite{lecun1998gradient} and USPS~\cite{lecun1990handwritten} are the two representative handwritten digit data sets widely used for evaluating the performance of machine learning and pattern recognition. For the transfer learning task, $2000$ $28 * 28$ images of the MNIST and $1800$ $16 * 16$ images of the USPS are selected as the source and target domain data sets. In the data preprocessing, all the images are linearly rescaled to $16 * 16$ meaning that the gray values of each image are represented by a $256$-dimensional vector. The MNIST and USPS share the same feature space but are generated from different distributions.

Concretely, the quantum circuit adopted by the VQCORAL has an $8$-qubit $16$-layer structure. The simulation results of the NA, the classical CORAL and the VQCORAL applied to the MNIST $\rightarrow$ USPS task and the USPS $\rightarrow$ MNIST task are presented in Table.~\ref{tab:MNIST_USPS}.

\begin{table}[htb]
	\caption{\label{tab:MNIST_USPS}%
	Accuracies of the NA, the classical CORAL, and the VQCORAL applied on the MNIST and USPS handwritten digit data sets.
	}
	\begin{ruledtabular}
		\begin{tabular}{ccc}
			\textrm{}&
			\textrm{$\textrm{MNIST} \rightarrow \textrm{USPS}$}&
			\textrm{$\textrm{USPS} \rightarrow \textrm{MNIST}$}\\
			\colrule
			NA & 64.4\% & 35.9\%\\
			Classical CORAL & 65.6\% & 46.9\%\\
			VQCORAL & \textbf{65.6\%} & \textbf{44.5\%}\\
		\end{tabular}
	\end{ruledtabular}
\end{table}

According to Table.~\ref{tab:MNIST_USPS}, both the classical CORAL and the VQCORAL show better performance than the NA meaning that the CORAL is helpful in accomplishing transfer learning tasks. In addition, the VQCORAL can achieve a comparable accuracy, namely $65.6 \%$, as the classical CORAL in the MNIST $\rightarrow$ USPS task. Although in the USPS $\rightarrow$ MNIST task, the accuracy of the VQCORAL is $44.5 \%$ which is not as good as the classical CORAL, the VQCORAL still exhibits better performance than the NA. We believe that the VQCORAL can achieve at least the comparable accuracy to the classical CORAL by further optimizing the design of the quantum circuit.  

\section{Discussions}
\label{sec:discussions}
In this paper, we propose two quantum versions of the CORAL, one of the most representative domain adaptation algorithms. On the one hand, the QCORAL implemented by the QBLAS can be performed on a universal quantum computer with exponential speedup in the dimension and number of the given data. On the other hand, the VQCORAL can be performed on the near term quantum devices with low circuit depth. Specifically, the VQCORAL can be implemented in two different perspectives. From an intuitive perspective, the VQCORAL can be realized directly by an end-to-end hierarchical structure. In addition, the source domain data can be decorrelated and aligned to the target domain data by successively applying the variational quantum covariance matrix square root solver and the variational matrix multiplication operations. To evaluate the feasibility and efficiency of our work, we design three different types of numerical experiments, namely the synthetic data, the synthetic-Iris data and the handwritten digit data. According to the simulation results, the VQCORAL presented in this paper can achieve at least comparable or even better performance than the classical CORAL.

However, some open questions need further study. First of all, the QBLAS-based CORAL requires a high-depth quantum circuit and fully coherent evolution which are actually prohibited in experiment at present. In addition, although the VQCORAL algorithm can be realized with limited quantum resources, the performance of the variational algorithm actually depends largely on the specific design of the parameterized circuits. Hence, it is well worth exploring how to design quantum circuits specifically to achieve optimal performance. Although some further exploration is required, it is demonstrated that quantum techniques can make a contribution to the field of domain adaptation.

\begin{acknowledgements}
	This work is supported by the National Key R\&D Program of China, Grant No. 2018YFA0306703.
\end{acknowledgements}

\nocite{*}

\bibliography{QCORAL.bib}% Produces the bibliography via BibTeX.
\bibliographystyle{apsrev4-2}

\end{document}